\documentclass[11pt,a4paper]{article}
\pdfoutput=1

\usepackage{bm}
\usepackage{amsmath,amssymb}
\usepackage{cite}

\usepackage{graphicx}
\usepackage{color}
\usepackage{upgreek}

\usepackage{hyperref} 
\hypersetup{colorlinks=true, citecolor=blue, filecolor=black, linkcolor=blue, urlcolor=blue, pdfpagemode=UseNone}

\numberwithin{figure}{section}
\numberwithin{equation}{section}

\newcommand{\be}{\begin{equation}}
\newcommand{\ee}{\end{equation}}
\newcommand{\bea}{\begin{eqnarray}}
\newcommand{\eea}{\end{eqnarray}}
\def\beal#1\eeal{\begin{align}#1\end{align}}   
\def\besp#1\eesp{\begin{multline}#1\end{multline}} 

\newcommand{\notes}[1]{}

\newcommand{\cL}{\mathcal{L}}

\newcommand{\Op}{\mathcal{O}}
\newcommand{\ph}{\varphi}
\newcommand{\vp}{\varphi}
\newcommand{\tp}{\tilde{\vp}}
\newcommand{\vpi}{\uppi}

\newcommand{\tg}{\tilde{g}}

\newcommand{\Ls}{\Lambda_\sigma}
\newcommand{\p}{\mathrm{p}} 
\newcommand{\dd}[2]{\delta_{\!\phantom{(} #1}^{\!(#2)}\!(\ph)}
\newcommand{\ddp}[3]{\delta_{\!\phantom{(} #1}^{\!(#2)}\!(#3)}

\newcommand{\Lm}[1]{\mathfrak{L}_{#1}}
\newcommand{\Ll}{\mathfrak{L}}
\newcommand{\Lmm}{\Lm-}

\newcommand{\ff}{\mathfrak{f}}

\newcommand{\htot}{H}

\newcommand{\prop}{\triangle}

\newcommand{\hs}{\hat{s}}

\newcommand\ie{\textit{i.e.}\ }
\newcommand\eg{\textit{e.g.}\ }
\newcommand\cf{\textit{cf.}\ }

\newcommand{\aka}{{a.k.a.}\ }

\newcommand{\viz}{{\it viz.}\ }
\newcommand{\half}{\tfrac{1}{2}}

\newcommand{\eps}{\varepsilon}

\newcommand{\morri}{Morris:2018mhd}
\newcommand{\morrii}{Morris:2018axr}
\newcommand{\yuji}{Igarashi:2019gkm}
\newcommand{\nn}{\nonumber}
\newcommand{\propH}{\prop_\Lambda}

\newcommand{\cu}[1]{\!#1\!}
\newcommand{\gh}[3]{$[#1,#2,#3]$}
\newcommand{\cG}{\check{\Gamma}}

\begin{document}

\begin{titlepage}

\begin{center}
{\huge \bf The continuum limit of quantum gravity at first order in perturbation theory}


\end{center}
\vskip1cm


\begin{center}
{\bf Alex Mitchell and Tim R. Morris}
\end{center}

\begin{center}
{\it STAG Research Centre \& Department of Physics and Astronomy,\\  University of Southampton,
Highfield, Southampton, SO17 1BJ, U.K.}\\
\vspace*{0.3cm}
{\tt  A.Mitchell-Lister@soton.ac.uk, T.R.Morris@soton.ac.uk}
\end{center}

\abstract{The Wilsonian renormalization group (RG) properties of the conformal factor of the metric are profoundly altered by the fact that it has a wrong-sign kinetic term. The result is a novel perturbative continuum limit for quantum gravity, which is however non-perturbative in $\hbar$. The ultraviolet part of the renormalized trajectory lies outside the diffeomorphism invariant subspace, entering this subspace only in the infrared, below a dynamically generated amplitude suppression scale. Interactions are dressed with coefficient functions of the conformal factor, their form being determined by the RG. In the ultraviolet, the coefficient functions are parametrised by an infinite number of underlying couplings. Choosing these couplings appropriately, the coefficient functions trivialise on entering the diffeomorphism invariant subspace. Here, dynamically generated effective diffeomorphism couplings emerge, including Newton's constant. In terms of the Legendre effective action, we establish the continuum limit to first order, characterising the most general form of such coefficient functions so as to verify universality.}


\end{titlepage}

\tableofcontents


\section{Introduction}
\label{sec:intro}

\begin{figure}[ht]
\centering
\includegraphics[scale=0.3]{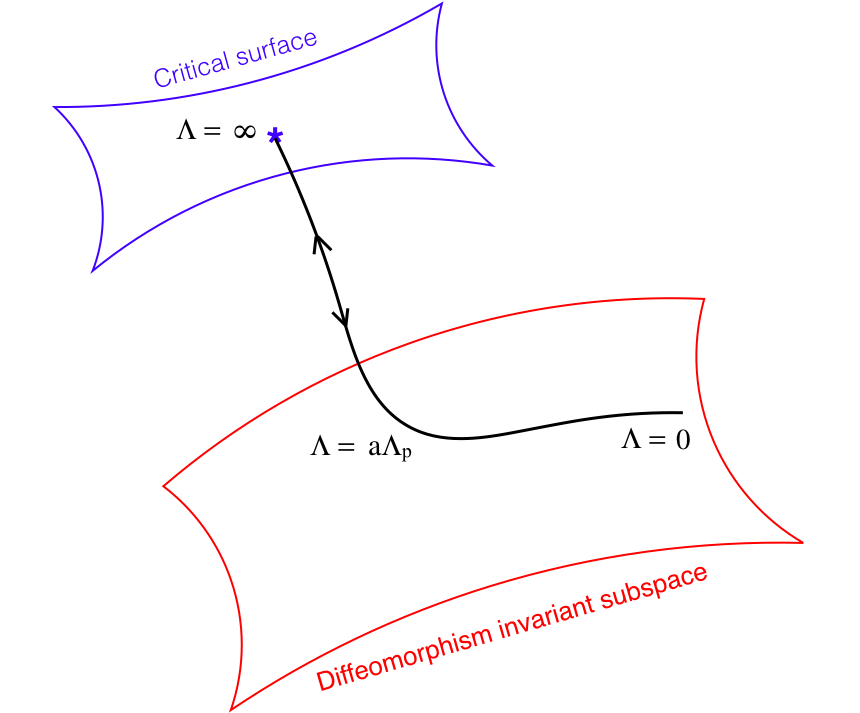}
\caption{The continuum limit is described by a renormalized trajectory that shoots out of the Gaussian fixed point (free gravitons) along relevant directions that cannot respect diffeomorphism invariance for $\Lambda>a\Lambda_\p$, where $\Lambda_\p$ is a characteristic of the renormalized trajectory and is called the \emph{amplitude suppression scale} (or \emph{amplitude decay scale}), and $a$ is a non-universal number. By appropriate choice of the \emph{underlying} couplings $g^\sigma_n$, diffeomorphism invariance is then recovered at scales $\Lambda,\vp\ll\Lambda_\p$ where also we recover an expansion in the \emph{effective} coupling  $\kappa\sim\sqrt{G}$.}
\label{fig:flow}
\end{figure}

In this paper we develop further  the perturbative continuum limit of quantum gravity begun in refs. \cite{\morri,Kellett:2018loq,Morris:2018upm,\morrii}. 
The theory is perturbative in $\kappa\sim \sqrt{G}$, the natural coupling constant (where $G$ is Newton's coupling), but non-perturbative in $\hbar$. It is the logical consequence of combining the Wilsonian RG (renormalization group) with the action for free gravitons, while respecting the wrong-sign kinetic term that then naturally appears in the conformal sector. Although this  renders the partition function meaningless without further reworking \cite{Gibbons:1978ac}, the Wilsonian RG remains well defined and provides us with an alternative and actually more powerful route to defining the quantum field theory. As such it then has all the usual desired properties (locality, microcausality, unitarity, gauge invariance \textit{etc.}) built in.
Nevertheless what we are led to is something conceptually different from all other approaches to quantum gravity, and indeed a construction crucially different from all other constructions of quantum field theories.

The basic structure of this continuum limit is illustrated in fig. \ref{fig:flow}, where we sketch the `theory space' of effective actions. In order to implement the Wilsonian RG structure one introduces a physical cutoff $\Lambda$ which sets the scale down to which modes are integrated out and allows us to define the Wilsonian effective action at this scale. Na\"\i vely one thinks of this cutoff as breaking the diffeomorphism invariance. However the Slavnov-Taylor identities get replaced by modified Slavnov-Taylor identities (mST), ``$\Sigma=0$'',  that reduce to the usual ones in the limit that we integrate out all the modes \cite{Ellwanger:1994iz,\yuji}. This limit is $\Lambda\to0$, which is the limit we need, in order to compute the desired physical observables. Now, because the mST are compatible with the flow equation, if the effective action enters this ``diffeomorphism invariant'' theory subspace at some (finite) scale $\Lambda$, \ie such that $\Sigma$ vanishes there, it never leaves this subspace, and physical quantities are then guaranteed to be diffeomorphism invariant.

So far so standard. However, what we find is that the ultraviolet fixed point which supports the continuum limit (at $\Lambda\to\infty$ and which for us is just the Gaussian fixed point, hence perturbatively describable), is located \emph{outside} the diffeomorphism invariant subspace, so that interactions constructed from the relevant operators cannot be made to satisfy $\Sigma=0$ there. Instead, by appropriate choice of the associated couplings $g^\sigma_n$, the renormalized trajectory joins the diffeomorphism invariant subspace in the limit as $\Lambda\ll\Lambda_\p$ (and also the conformal mode must have amplitude $\vp\ll\Lambda_\p$) where $\Lambda_\p$ is a dynamically generated scale determined by the underlying couplings, called the \emph{amplitude suppression scale} \cite{\morri}. Equivalently, in the limit in which this new scale  $\Lambda_\p\to\infty$, we have $\Sigma\to0$ and diffeomorphism invariance is recovered. Here we recover Newton's constant as another dynamically generated scale determined by these underlying couplings, and as we'll see also the cosmological constant. 

Let us emphasise  that this structure  follows inevitably from imposing the principles of the Wilsonian RG about the Gaussian fixed point, while taking seriously the consequences of the wrong-sign kinetic term in the conformal sector and requiring that in physical amplitudes we recover diffeomorphism invariance \cite{\morrii}. It is therefore well grounded and indeed thus may not seem so different from the usual picture. However all other quantum field theories have Wilsonian RG flows that \emph{can} be defined within the gauge invariant subspace. For example for (non-Abelian) gauge theories the continuum solution can be chosen to respect the corresponding $\Sigma=0$ identities at all scales, \eg \cite{\yuji}, in fact
the gauge invariance can even be manifestly respected through \eg lattice regularisation \cite{Wilson:1974sk} or directly in the continuum (\eg \cite{Morris:1999px,Morris:2000fs,Arnone:2005fb}). 

In \emph{all other approaches} to quantum gravity, it  has  been  assumed that the Wilsonian RG properties defining the continuum limit, and the diffeomorphism gauge invariance, can coexist in the same region of the renormalized trajectory. This tacit assumption for example lies behind intuitive arguments against the existence of an ultraviolet fixed point in quantum gravity, based on black hole entropy considerations \cite{Aharony:1998tt,Shomer:2007vq}. We see that these arguments are actually inapplicable in this case.\footnote{It has been argued that they do not apply in the asymptotic safety scenario either, but for different reasons \cite{Falls_2014}.} To put it pithily, such tensions in quantum gravity are resolved since a crucial element of quantum gravity is constructed \emph{off} space-time. This is to be contrasted with classical General Relativity which is a construction \emph{of} space-time, and with normal quantum field theories which are constructed \emph{on} space-time.


There are other important properties,  which are key to a complete understanding of  fig. \ref{fig:flow}, especially the fact that the operators are not those of the usual expansion but non-polynomial in $\vp$, that infinitely many of these are relevant, that the expansion in terms of these operators actually only makes sense at scales above $a\Lambda_\p$, and that flows in the conformal sector go in the reverse direction (from infrared to ultraviolet). In ref. \cite{\morrii} we highlighted how these novelties lead to differences that need careful treatment. These include differences in limiting procedures, in particular
ultraviolet divergences are now absorbed by the underlying couplings, while at low scales outside the diffeomorphism invariant subspace new infrared divergences appear \cite{\morri}.  For these reasons, in this paper we develop further the properties at first order, and provide a tight characterisation of the most general form of the continuum limit at this order as needed for the higher order computations.

The first order continuum limit was formulated in ref. \cite{\morrii} in terms of the Wilsonian effective action and a regularised Quantum Master Equation. Although this allows for an elegant analysis since the latter effectively leaves  BRST invariance unmodified, in sec. \ref{sec:prelim} we switch to an equivalent \cite{Morris:1993,Morris:2015oca} description  in terms of the infrared regulated Legendre effective action \cite{Nicoll1977, Wetterich:1992, Morris:1993} and the mST \cite{Ellwanger:1994iz,\yuji}. Although more cumbersome, this then gives us direct access to the one-particle irreducible amplitudes in the physical limit, and leads to useful simplifications at higher orders \cite{\morri,secondconf,second}. Furthermore it can still be solved in terms of the total free quantum BRST charge $\hs_0$ \cite{\yuji} that naturally incorporates a regularised Batalin-Vilkovisky measure operator $\Delta$ \cite{Batalin:1981jr,Batalin:1984jr}.  

In secs. \ref{sec:prelim} and \ref{sec:cohomology} we review the development of this free BRST algebra and how computations can be couched in minimal gauge invariant basis \cite{\yuji}. Then in sec. \ref{sec:nontrivialcohomology}, we choose the first order non-trivial quantum BRST cohomology representative on which to build the continuum limit to first order (in the new quantisation these two are not the same). In order to simplify the higher order computations \cite{second}, we choose one that corresponds to expressing diffeomorphism invariance as a Lie derivative, and demonstrate that this differs from the previous choice \cite{\morrii,Boulanger:2000rq} by an $\hs_0$-exact piece, such that the regularised measure term $\Delta$ provides a contribution crucial for consistency.

In sec. \ref{sec:newquantisation} we review how the wrong sign kinetic term in the conformal sector profoundly alters RG properties that are central to defining the continuum limit, however framing the discussion now in terms of the Legendre effective action. In particular we recall how this leads to all interactions $\sigma$ being dressed with a coefficient function $f_\Lambda^\sigma(\vp)$. This latter is parametrised by the underlying couplings $g^\sigma_n$. At the linearised level only those couplings of non-negative mass dimension must be non-vanishing. Here we work with the most general such coefficient functions that are consistent with the RG properties as determined by the flow equation, and such that the renormalized trajectory enters the diffeomorphism invariant subspace as sketched in fig. \ref{fig:flow}. We do so in order to verify the universality of this continuum limit, here at first order, and later at higher orders \cite{secondconf,second}. We tighten and further develop the arguments from refs. \cite{\morri,\morrii}, that show how the RG properties  determine the form of the dressed interactions and their coefficient functions. In doing so, we demonstrate once again that these results follow inevitably from combining the Wilsonian RG and the Gaussian fixed point action for free gravitons, after taking seriously the consequences of the resulting wrong-sign kinetic term in the conformal sector. In particular we give closed expressions for the tadpole corrections appearing in the dressed interactions, prove that there exists a dynamically generated amplitude suppression scale $\Lambda_\sigma$ that determines the large $\vp$ behaviour of each coefficient function $f_\Lambda^\sigma(\vp)$ for all $\Lambda\cu\ge0$ and prove that $f_\Lambda^\sigma(\vp)$ itself is determined uniquely by its physical limit. Finally we show that these are given in conjugate momentum space by an entire function $\ff^\sigma(\vpi)$ whose Taylor expansion coefficients are the underlying couplings $g^\sigma_n$.

In sec. \ref{sec:newquantisation} and sec. \ref{sec:largeamp} we show that in turn the amplitude suppression scale characterises the asymptotic behaviour of the underlying couplings $g^\sigma_n$ at large $n$. In sec. \ref{sec:largeamp} we define what it means for the coefficient functions to \emph{trivialise} in the the large $\Lambda_\sigma$ limit. From ref. \cite{\morrii} we know that the underlying couplings must be chosen so that this trivialisation happens, in order to enter the diffeomorphism invariant subspace at the linearised level. In the simplest case this means that the coefficient function must tend to a constant in this limit; more generally we show that it must tend to a Hermite polynomial of degree $\alpha$, whose functional form is then fixed. 

In sec. \ref{sec:relations} we show how to derive new solutions for coefficient functions from a given one, and derive formulae for their underlying couplings, either by multiplying the physical coefficient by a power of $\vp$ or by differentiating with respect to $\vp$. These tricks prove useful later. 

Then in sec. \ref{sec:general} we characterise the most general form of coefficient functions that trivialise in the large $\Lambda_\sigma$ limit. This is most efficiently expressed in terms of  their Fourier transform. In particular we show that $\ff^\sigma(\vpi)$ must tend to (the $\alpha^\text{th}$ derivative of) a Dirac $\delta$-function. We make two powerful simplifying assumptions which still leave us with an infinite dimensional function space of solutions flexible enough to encompass the higher order computations. Firstly we specialise to coefficient functions that have definite parity (are even or odd functions). Secondly we insist that at the linearised level the coefficient functions contain only one amplitude suppression scale.\footnote{However in app. \ref{app:spectrum}, we also develop their properties when there is a spectrum of amplitude suppression scales.}  Putting all these properties together, allows us to give a complete characterisation of $\ff^\sigma(\vpi)$ in terms of its large and small $\vpi$ behaviour, its normalisation, and limiting behaviour of key integrals at large $\Lambda_\sigma$. In particular we use this to characterise the approach to the trivialisation limit. In sec. \ref{sec:examples} we verify all these general properties on a series of instructive examples. 

Finally in sec. \ref{sec:first} we construct a very general continuum limit to first order, and verify that its renormalization group trajectory fulfills the properties sketched in fig. \ref{fig:flow}. We finish the paper in sec. \ref{sec:discussion} by discussing the meaning and implications of this construction and its relation to other approaches.

\section{Legendre effective action, mST, and quantum gravity}
\label{sec:prelim}

We begin by briefly recalling some key steps from refs. \cite{\morri,\morrii,\yuji}. This will also serve to set out our choice of notation and formulation for this paper.
In ref. \cite{\morrii}, we worked with the continuum Wilsonian effective action. Here we will work directly with the renormalized infrared cutoff Legendre effective action $\Gamma$, which is also in fact the one-particle irreducible part of the continuum Wilsonian effective action \cite{Morris:1993}. However it will mean that BRST invariance is no longer expressed as unbroken through the  Quantum Master Equation but rather through modified Slavnov-Taylor identities (mST) \cite{Ellwanger:1994iz,\yuji}, so that we recover (off-shell) nilpotency at the interacting level, only in the limit $\Lambda\cu\to0$. The free charges are still nilpotent however, and it is their cohomology that is central to solving for the effective action \cite{\yuji}.
In any case the loss of some elegance is outweighed by the advantages: the simplification that comes from not computing also the one-particle reducible parts and especially the fact that the limit then gives us direct access to the physical amplitudes:
\be 
\label{physical}
\Gamma_\text{phys} = \lim_{\Lambda\to0} \Gamma\,.
\ee
The flow equation for the interacting part thus takes the form \cite{Nicoll1977, Wetterich:1992, Morris:1993} (see also  \cite{Weinberg:1976xy,Morris:2015oca,Bonini:1992vh,Ellwanger1994a,Morgan1991}):
\be 
\label{flow}
\dot{\Gamma}_I = -\half\, \text{Str}\left( \dot{\prop}_\Lambda\propH^{-1} \left[1+\propH \Gamma^{(2)}_I \right]^{-1}\right) 
\,,
\ee
where the over-dot is $\partial_t =-\Lambda \partial_\Lambda$. The BRST invariance is  expressed through the mST  \cite{Ellwanger:1994iz,\yuji}:
\be 
\label{mST}
\Sigma := \half (\Gamma,\Gamma) - \text{Tr}\left( \!C^\Lambda\,  \Gamma^{(2)}_{I*} \left[1+\propH\Gamma^{(2)}_I\right]^{-1}\right) = 0\,.
\ee
These equations are both ultraviolet (UV) and infrared (IR) finite thanks to the presence of the UV cutoff function $C^\Lambda(p)\equiv C(p^2/\Lambda^2)$ which, since it is multiplicative, satisfies $C(0)=1$, and its associated IR cutoff
$C_\Lambda= 1-C^\Lambda$, which appears in the IR regulated propagators as $\propH^{AB} = C_\Lambda\prop^{AB}$. The cutoff function is chosen so that $C(p^2/\Lambda^2)\cu\to0$ sufficiently fast as $p^2/\Lambda^2\cu\to\infty$ to ensure that all momentum integrals are indeed UV regulated (faster than power fall off is necessary and sufficient). It is also required to be smooth (differentiable to all orders), corresponding to a local Kadanoff blocking. It thus permits for $\Lambda\cu>0$, a quasi-local solution for $\Gamma_I$, namely one that has a space-time derivative expansion to all orders. We insist on this: 
it is equivalent to imposing locality on a bare action.
 
The two equations are compatible: 
if $\Sigma=0$ at some generic scale $\Lambda$, it remains so on further evolution, in particular as $\Lambda\to0$. The second term in the mST \eqref{mST}
is a quantum modification due to the cutoff $\Lambda\cu>0$. At non-exceptional momenta (\ie such that no internal particle in a vertex can go on shell) it remains IR finite, and thus vanishes as $\Lambda\to0$, thanks to the UV regularisation. We are then left with just the Zinn-Justin equation $\half(\Gamma,\Gamma)=0$ \cite{ZinnJustin:1974mc,ZinnJustin:1975wb}, which gives us the standard realisation of quantum BRST invariance through the Slavnov-Taylor identities for the corresponding vertices.

In the above equations we have introduced Str$\,\mathcal{M} = (-)^A\, \mathcal{M}^{A}_{\ \,A}$ and Tr$\,\mathcal{M} = \mathcal{M}^{A}_{\ \,A}$, and set
\be 
\label{Hessian}
\Gamma^{(2)}_{I\ AB} = \frac{\partial_l}{\partial\Phi^A}\frac{\partial_r}{\partial\Phi^B}\Gamma_I\,,\qquad \
\left(\Gamma^{(2)}_{I*}\right)^{A}_{\ \ B} 
\,=\,  \frac{\partial_l}{\partial\Phi^*_A}\frac{\partial_r}{\partial\Phi^B}\Gamma_I\,,
\ee 
Here $\Phi$ and $\Phi^*$ are the collective notation for the classical fields and antifields (sources of BRST transformations) respectively,
while $\Gamma$ is the ``effective average action'' \cite{Wetterich:1992} part of the infrared cutoff Legendre effective action \cite{\yuji,Morris:1993}:
\be 
\label{LegendreEffAct}
\Gamma^{tot} = \Gamma + \half \Phi^A \mathcal{R}_{AB} \Phi^B\,,  \qquad \prop^{-1}_{\Lambda\, AB} = \prop^{-1}_{AB} + \mathcal{R}_{AB}\,,
\ee
where $\mathcal{R}_{AB}$ is the infrared cutoff expressed in additive form.
$\Gamma$ is expressed in terms of a free part, $\Gamma_0$, which includes the free BRST transformations, plus the interaction part $\Gamma_I[\Phi,\Phi^*]$:
\be 
\label{Gamma}
\Gamma = \Gamma_0 + \Gamma_I\,,\qquad \Gamma_0 =
\half\, \Phi^A \prop^{-1}_{AB}\Phi^B -  (Q_0\Phi^A) \Phi^*_A \,.  
\ee
Note that the free part carries no regularisation.
The antibracket in the mST is similarly expressed without regularisation. For arbitrary functionals of the classical (anti)fields, $\Xi[\Phi,\Phi^*]$ and $\Upsilon[\Phi,\Phi^*]$, it is given by
\be 
\label{QMEbitsPhi}
(\Xi,\Upsilon) = \frac{\partial_r\Xi}{\partial\Phi^A}\,\frac{\partial_l\Upsilon}{\partial\Phi^*_A}-\frac{\partial_r\Xi}{\partial\Phi^*_A}\,\frac{\partial_l\Upsilon}{\partial\Phi^A}\,.
\ee 
Notice that in $\Gamma_0$ we have chosen left-acting BRST transformations \cite{\morrii} (see also app. A2 of \cite{\yuji}) so that the free BRST transformation is given by the first of the following equations:
\be 
\label{charges}
Q_0\, \Phi^A := (\Gamma_0,\Phi^A)\,,\qquad Q^-_0\Phi^*_A := (\Gamma_0,\Phi^*_A)\,.
\ee
Here we have taken the opportunity also to define the free Koszul-Tate operator $Q^-_0$.

We will be interested in expanding $\Gamma_I$ perturbatively in its interactions, assuming the existence of an appropriate small parameter $\epsilon$:
\be 
\label{expansion}
\Gamma_I = \sum_{n=1}^\infty\Gamma_n\,{\epsilon^n}/{n!} \,.
\ee
Importantly, in the quantisation established in \cite{\morri,\morrii}, we need however  to work non-perturbatively in $\hbar$, so there will be no loop expansion. In the above, $\epsilon$ is a formal perturbation-order counting parameter, which we set to $\epsilon=1$ at the end. The actual small physical parameter,
\be 
\label{kappa}
\kappa=\sqrt{32\pi G} 
\ee
(where $G$ is Newton's gravitational constant)
will properly make its appearance in the theory only in sec. \ref{sec:first}, where it arises as a collective effect of all the underlying couplings.

At first order the flow equation \eqref{flow} and mST \eqref{mST} become
\beal 
\label{flowone}
\dot{\Gamma}_1 &=  \half\, \text{Str}\, \dot{\prop}_\Lambda \Gamma^{(2)}_1 \,, \\
\label{mSTone}
0 &=  (\Gamma_0,\Gamma_1) - \text{Tr}\left( \!C^\Lambda\,  \Gamma^{(2)}_{1*} \right) = (Q_0+Q^-_0-\Delta) \Gamma_1 =: \hs_0\, \Gamma_1\,,
\eeal
where the first equation is the flow equation satisfied by eigenoperators: their RG time derivative is given by the action of the tadpole operator \cite{\morrii}.
In the second equation we recognise that we recover the  Batalin-Vilkovisky measure operator \cite{Batalin:1981jr,Batalin:1984jr}:
\be 
\label{Delta}
\Delta  = (-)^A \frac{\partial_l}{\partial\Phi^A}\,C^\Lambda\frac{\partial_l}{\partial\Phi^*_A}\,,
\ee
UV regulated as in refs. \cite{\morrii,\yuji},
and we have defined the corresponding full free quantum BRST charge $\hs_0$.
Note that $\Delta$ thus generates $\Lambda$-dependent tadpole integral corrections to the full free classical BRST transformations. Thanks to compatibility, these corrections are as required in order to find simultaneous solutions of the linearised flow equation \eqref{flowone} and linearised mST \eqref{mSTone}.
Indeed, as shown in \cite{\morrii}, the $\hs_0$-cohomology can then be defined within the space spanned by the eigenoperators with constant coefficients (\aka couplings).

In this paper, any explicit expression for an action functional should be understood as integrated over four flat Euclidean spacetime dimensions and determined only up to integration by parts. 
As we will explain shortly, we can in effect work in minimal gauge invariant basis \cite{\yuji}
where 
\be 
\label{Gzero}
\Gamma_0 = \half \left(\partial_\lambda \htot_{\mu\nu}\right)^2 -2 \left(\partial_\lambda \ph\right)^2 - \left(\partial^\mu \htot_{\mu\nu}\right)^2 +2\,\partial^\alpha\! \ph\, \partial^\beta \htot_{\alpha\beta}\  - 2\,\partial_\mu c_\nu H^*_{\mu\nu}
\ee
is the action for free graviton fields $H_{\mu\nu}$, plus the fermionic antifield $H^*_{\mu\nu}$ source term for
\be 
\label{QH}
Q_0 H_{\mu\nu} = \partial_\mu c_\nu+\partial_\nu c_\mu\,,
\ee
the only non-vanishing free linearised BRST transformation in this basis,  this matching the general form \eqref{Gamma}, $c_\mu$ being the (fermionic) ghost fields. Contraction is with the flat metric $\delta_{\mu\nu}$, and we write $\ph = \half\,H_{\mu\mu}$.
Since raising an index makes no difference we will usually leave all indices as subscripts. 

We note in passing that the free action \eqref{Gzero} is also the action one gets from the Einstein-Hilbert Lagrangian 
\be 
\label{EH}
\cL_{EH} = -2\sqrt{g}R/\kappa^2\,,
\ee
if one expands the metric as
\be 
\label{gH}
g_{\mu\nu} = \delta_{\mu\nu} +\kappa H_{\mu\nu}\,.
\ee
Similarly the  free BRST invariance \eqref{QH} follows from expanding diffeomorphisms (regarding $\kappa c^\mu$ as the small diffeomorphism).

The only extra (anti)field we will need is the bosonic $c^*_\mu$, 
the source for BRST transformations of $c_\mu$ that will appear at the interacting level. 
From the general definition \eqref{charges} and the free action \eqref{Gzero}, the non-vanishing free Kozsul-Tate differentials are:
\be 
\label{KTHc}
Q^-_0 H^*_{\mu\nu} = -2G^{(1)}_{\mu\nu}\,,\qquad Q^-_0 c^*_\nu = -2 \partial_\mu H^*_{\mu\nu}\,,
\ee
where $G^{(1)}_{\mu\nu}$ is the linearised Einstein tensor:
\be 
\label{Gmunu}
G^{(1)}_{\mu\nu} = -R^{(1)}_{\mu\nu}+\half R^{(1)}\delta_{\mu\nu} = \half\, \Box\, H_{\mu\nu} -\delta_{\mu\nu}\Box\,\vp+\partial^2_{\mu\nu}\vp+\half \delta_{\mu\nu}\partial^2_{\alpha\beta} H_{\alpha\beta}-\partial_{(\mu}\partial^\alpha H_{\nu) \alpha}\,,
\ee
the linearised curvatures being\footnote{defining symmetrisation as: $t_{(\mu\nu)} = \half (t_{\mu\nu}+t_{\nu\mu})$, and antisymmetrisation as $t_{[\mu\nu]} = \half (t_{\mu\nu}-t_{\nu\mu})$.}
\be 
\label{curvatures}
R^{(1)}_{\mu\alpha\nu\beta} = -2 \partial_{[\mu|\,}\partial_{[\nu} H_{\beta]\,|\alpha]}\,,\  R^{(1)}_{\mu\nu} = -\partial^2_{\mu\nu}\vp+\partial_{(\mu}\partial^\alpha H_{\nu) \alpha}-\half\, \Box\, H_{\mu\nu}\,,\  R^{(1)} = \partial^2_{\alpha\beta}H_{\alpha\beta}-2\,\Box\,\vp\,.
\ee
It is evident that the Koszul-Tate transformations \eqref{KTHc} are invariances of the free action \eqref{Gzero}, the former by the linearised Bianchi identity and the latter trivially so.

In order to derive the propagators, which are used in both the flow equation \eqref{flow} and the mST \eqref{mST}, we need to introduce gauge fixing.
To do this we first extend to the non-minimal basis by adding the bosonic auxiliary field $b_\mu$ that allows off-shell BRST invariance, and $\bar{c}^*_\mu$ which sources BRST transformations of the antighost $\bar{c}_\mu$. Then the free effective action is written as \cite{\morrii}:
\be 
\Gamma_0 |_\text{gi} = \Gamma_0 +\frac{1}{2\alpha}b^2_\mu -i b_\mu \bar{c}^*_\mu\,,
\ee
where $\alpha$ is our gauge fixing parameter. Gauge fixing is implemented by a finite quantum canonical transformation \cite{Siegel:1989ip,Gomis:1994he} that takes us to gauge fixed basis 
$
\Phi^*_A |_\text{gf} = \Phi^*_A |_\text{gi} + \partial^r_A \Psi
$,
where $\Psi$ is the gauge fixing fermion. Choosing
$ 
\Psi = \bar{c}_\mu F_\mu
$,
where $F_\mu = \partial_\nu H_{\nu\mu} -\partial_\mu\vp$ is  De Donder gauge,
the canonical transformation only changes $\bar{c}^*_\mu\,|_\text{gi}  = \bar{c}^*_\mu \,|_\text{gf} - F_\mu$ and:\footnote{defining vector contraction as $u\!\cdot\! v = u_\mu v_\mu$.}
\be
\label{gaugeFixed}
H^*_{\mu\nu} \,|_\text{gi} = H^*_{\mu\nu} \,|_\text{gf} +\partial_{(\mu} \bar{c}_{\nu)} -\half\,\delta_{\mu\nu}\, \partial\!\cdot\! \bar{c}\,.
\ee
The free action in gauge fixed basis is therefore:
\be 
\Gamma_0 |_\text{gf} = \Gamma_0 -\bar{c}_\mu\Box\, c_\mu -ib_\mu F_\mu +\frac{1}{2\alpha}b^2_\mu -i b_\mu \bar{c}^*_\mu\,.
\ee
It has kinetic operators that can be inverted.
The $H_{\mu\nu}$ propagator simplifies in  ``Feynman gauge'' $\alpha=2$, which as in ref. \cite{\morrii} we set from now on. Splitting $H_{\mu\nu}$ into its SO$(4)$ irreducible parts,
\be 
\label{h}
H_{\mu\nu} = h_{\mu\nu} +\half\, \ph\, \delta_{\mu\nu}
\ee
(thus $h_\mu^{\ \mu} = 0$ is traceless), in this gauge the two parts decouple.
The propagators we need are
\beal
\label{hh}
\langle h_{\mu\nu}(p)\,h_{\alpha\beta}(-p)\rangle &= \frac{\delta_{\mu(\alpha}\delta_{\beta)\nu}-\frac14\delta_{\mu\nu}\delta_{\alpha\beta}}{p^2} 
\,,\\
\label{pp}
\langle \ph(p)\,\ph(-p)\rangle &=  -  \frac1{p^2}\,,\\
\label{cc}
\langle c_\mu(p)\, \bar{c}_\nu(-p)\rangle &= -\langle \bar{c}_\mu(p)\, c_\nu(-p) \rangle =  \delta_{\mu\nu}/{p^2}\,,
\eeal
where we have written \notes{477.7}
\be 
\label{defs}
\prop^{AB}  = \langle\Phi^A\,\Phi^B\rangle\,,\qquad
\Phi^A(x) = \int_p  \text{e}^{-i p\cdot x}\, \Phi^A(p)\,,\qquad
\int_p \equiv \int\!\! \frac{d^4p}{(2\pi)^4}\,.
\ee
Note that $h_{\mu\nu}$ propagates with the right sign, and that the numerator  is just the projector onto traceless tensors, while $\vp$ propagates with wrong sign. 

There is a propagator involving $b_\alpha$ \cite{\morrii} but it is not needed. Indeed,
we will later confirm that the first order interaction $\Gamma_1$ can be constructed just from the minimal set. Then in gauge fixed basis, $\Gamma_1$ still does not depend on $b_\mu$ or $\bar{c}^*_\mu$ and will depend on $\bar{c}_\mu$ only through the combination on the right hand side (RHS) of the transformation to gauge fixed basis \eqref{gaugeFixed}. By iteration, using the flow equation \eqref{flow}, these properties are inherited by all the higher order interactions $\Gamma_{n>1}$. Mapping back to gauge invariant basis using the equations above, we therefore see that $\Gamma_I$ will not depend on $b_\mu$, $\bar{c}^*_\mu$ or $\bar{c}_\mu$. This means in particular that the full $\Gamma_I$ remains in {minimal} gauge invariant basis. 

Therefore we can most simply express the calculation in this basis \cite{\yuji} as we will do from now on. What this means is that when we compute corrections from the flow equation \eqref{flow} or from the quantum correction part of the mST \eqref{mST}, we temporarily make the shift to gauge fixed basis using \eqref{gaugeFixed}, which in particular then allows corrections computed using the ghost propagator \eqref{cc}, after which we absorb the antighost by shifting back to minimal gauge invariant basis using the inverse of \eqref{gaugeFixed}. Notice that since the transformation is canonical, it has no effect on the antibracket part of the mST \eqref{mST} which thus can be computed whilst remaining in (minimal) gauge invariant basis.



\section{Free quantum BRST cohomology}
\label{sec:cohomology}

Following Henneaux \textit{et al}  \cite{Barnich:1993vg}, we can simplify finding solutions of the $\hs_0$-cohomology by splitting the problem up (\aka grading) by  antighost, \aka antifield, number. 
We thus have the weights $H^*_{\mu\nu}$ \gh{-1}12, $c^*_\mu$ \gh{-2}22, $H_{\mu\nu}$ \gh001 and $c_\mu$ \gh101,
where the first entry is the ghost number, the second entry the antighost/antifield number, and the final entry the mass dimension. (A full table of weights is given in ref. \cite{\morrii}.) Thus all parts of $\hs_0$ increase ghost number and mass dimension by one.  While ghost number and mass dimension are respected, antighost number is not, but it is chosen so that the free BRST charges have definite antighost number. We anticipated this with our labelling: $Q_0$  leaves antighost number unchanged, while $Q^-_0$ lowers it by one. Under this grading, the measure operator splits into two parts that lower antighost number by one or two respectively ($\Delta^-$ simplifies to this in minimal basis \cite{\morrii}):
\be 
\label{measure}
\Delta = \Delta^- + \Delta^=\,,\qquad\Delta^- = \frac{\partial}{\partial H_{\mu\nu}}C^\Lambda\frac{\partial_l}{\partial H^*_{\mu\nu}}  \,,\qquad \Delta^= = - \frac{\partial_l}{\partial c_\mu}C^\Lambda\frac{\partial}{\partial c^*_\mu}\,.
\ee
The point of this extra grading is that $\Gamma$ itself does not have definite antifield number but splits into parts of definite antifield number $n$: 
$
\Gamma=\sum_{n=0} \Gamma^n\,.
$
This means that an (integrated) operator $\Op = \sum_{m=0}^n \Op^m$ with some maximum antighost number $n$, that is annihilated by $\hs_0$, must satisfy the descent equations:
\be 
\label{descendants}
Q_0\, \Op^n = 0\,, \quad Q_0\, \Op^{n-1} = (\Delta^--Q^-_0)\, \Op^n\,,\quad Q_0\,\Op^{n-2} =  (\Delta^--Q^-_0) \,\Op^{n-1} + \Delta^=\,\Op^n\,, \cdots\,.
\ee
Starting with the top (left-most) equation, these are often easier to analyse than trying to work with $\hs_0\,\Op=0$ directly. Grading the square we also have the useful identities \cite{\morrii,\yuji}:\footnote{In ref. \cite{\morrii} we incorrectly assumed that the interacting BRST charges have definite antighost number (see footnote 10 of \cite{\yuji}), and thus that these identities hold in general, although we actually applied them only at the free level.} 
\beal
\label{nilpotents}
Q^2_0 =0\,,\ (Q^-_0)^2 = 0\,,\ &(\Delta^-)^2 = 0\,,\ (\Delta^=)^2 = 0\,,\nonumber\\
\{Q_0,Q^-_0\} =0\,,\ \{Q_0,\Delta^-\}=0\,,\ &\{Q^-_0,\Delta^=\}=0\,,\ \{\Delta^-,\Delta^=\}=0\,,\nonumber\\
\{Q^-_0,\Delta^-\}+& \{Q_0,\Delta^=\} = 0\,.
\eeal

\subsection{Non-trivial free quantum BRST cohomology representatives}
\label{sec:nontrivialcohomology}

As we will review in sec. \ref{sec:first}, our choice of non-trivial $\hs_0$-cohomology representative, $\cG_1$, will lead us to the solution for the first order interactions $\Gamma_1$. (The latter is not simply $\kappa\,\cG_1$ as it would be in standard quantisation \cite{\morrii}.) In order to get a theory that is consistent with unitarity and causality, we restrict $\cG_1$ to have a maximum of two space-time derivatives. Then $\cG_1$ must be a linear combination of a term involving space-time derivatives and a unique non-derivative piece:
\be 
\label{Gonecc}
\cG_1 = \cG^0_1 = \vp\,.
\ee
This latter is nothing but the $O(\kappa)$ part of $\sqrt{g}$, \ie what one gets from a classical cosmological constant term, using the first order expansion of the metric in terms of fluctuation field  \eqref{gH}.
The derivative part has a unique expression with maximum antighost number two \cite{\morrii,Boulanger:2000rq}, up to addition of $\hs_0$-exact pieces.
Previously we followed \cite{Boulanger:2000rq} in using the simplest form for this $\hs_0$-cohomology representative, 
which corresponds to treating $c_\mu$ as a covariant vector field \cite{\morrii}. At higher orders the formulae will simplify however if we treat $c^\mu$ as a contravariant vector field since diffeomorphisms can then be expressed through the Lie derivative and thus be independent of the metric. Then the maximum antighost number piece is
\be 
\label{Gonetwo}
\cG^2_1 = -\left( c^\mu\partial_\mu c^\nu\right) c^*_\nu = c_\mu\partial_\nu c_\mu c^*_\nu + Q_0\left( H_{\mu\nu} c_\mu c^*_\nu \right)\,,
\ee
where the first bracketed term is half the Lie bracket as required \cite{\morrii}, and in the second equality we use the free diffeomorphism BRST transformation \eqref{QH} to express it as the old choice plus a $Q_0$-exact piece (the first term on the $\hs_0$-exact addition \eqref{soeigenoperator}'s RHS). Now we could use the second expression and descend via the descendent equations \eqref{descendants} using the nilpotency relations \eqref{nilpotents}, but since we know that the expression is unique up to addition of $\hs_0$-exact pieces, we see immediately that our new choice must be
\be 
\label{newvsold}
\cG_1 = \cG_1 |_\text{old} + \hs_0\left( H_{\mu\nu} c_\mu c^*_\nu \right)\,,
\ee
up to possible further $\hs_0$-exact terms of lower antighost number. Using also the Koszul-Tate transformation  \eqref{KTHc} and the explicit formula for the Batalin-Vilkovisky measure \eqref{measure},
\be 
\label{soeigenoperator}
\hs_0\left( H_{\mu\nu} c_\mu c^*_\nu \right) = (\partial_\mu c_\nu + \partial_\nu c_\mu)\, c_\mu c^*_\nu +2H_{\mu\nu} c_\mu \partial_\alpha H^*_{\alpha\nu} + 2b\Lambda^4\vp\,,
\ee 
where we note that $\Delta^-$ trivially annihilates, but $\Delta^=$ yields a UV regulated quartically divergent contribution, $b$ being the non-universal number already introduced in refs. \cite{Morris:2018mhd,\morrii}:
\be 
\label{b}
b = \int\!\frac{d^4\tilde{p}}{(2\pi)^4}\, C(\tilde{p}^2)\,.
\ee
From the relation between the new and old choices \eqref{newvsold} and the antighost level one part of the $\hs_0$-exact addition \eqref{soeigenoperator}, we have
\be 
\label{Goneone}
\cG^1_1 = 2  c_\alpha \Gamma^{(1)\, \alpha}_{\phantom{(1)\,}\mu\nu}H^*_{\mu\nu} +2H_{\mu\nu} c_\mu \partial_\alpha H^*_{\alpha\nu} = - \left(c^\alpha \partial_\alpha H_{\mu\nu} + 2\, \partial_\mu c^\alpha H_{\alpha\nu}\right) H^*_{\mu\nu} \,,
\ee
where the previous choice involves the linearised connection
$ 
\Gamma^{(1)\, \alpha}_{\phantom{(1)\,}\mu\nu} = \half\left(\partial_\mu H_{\alpha\nu}\cu+\partial_\nu H_{\alpha\mu}\cu-\partial_\alpha H_{\mu\nu}\right)
$.
Integrating by parts we get the second expression, and we recognise that inside the brackets we already have the desired Lie derivative form. Combining it with the last term of the free action \eqref{Gzero}, the expression for the Lie derivative of the metric is given exactly, provided that the metric is taken to be exactly the definition \eqref{gH} in terms of $H_{\mu\nu}$. In other words at the classical level neither the first-order antighost level two part \eqref{Gonetwo} nor the first-order antighost level one part \eqref{Goneone} receives corrections at higher order in perturbation theory. 
Finally, from the relation between the new and old choices \eqref{newvsold} and the expression for $\hs_0$-exact addition \eqref{soeigenoperator}, we have
\beal
\label{Gonezero}
\cG^0_1\ &=\ \cG^0_1 |_\text{old}+2b\Lambda^4\vp\\
& =\ 2 \vp\partial_\beta H_{\beta\alpha}\partial_\alpha\vp 
-2\vp(\partial_\alpha\vp)^2
-2H_{\alpha\beta}\partial_\gamma H_{\gamma\alpha}\partial_\beta\vp
+2H_{\alpha\beta}\partial_\alpha\vp\partial_\beta\vp
-2H_{\beta\gamma}\partial_\gamma H_{\alpha\beta}\partial_\alpha\vp\nn \\
&\phantom{=\ } +\half\vp (\partial_\gamma H_{\alpha\beta})^2 
-\half H_{\gamma\delta}\partial_\gamma H_{\alpha\beta}\partial_\delta H_{\alpha\beta}
-H_{\beta\mu}\partial_\gamma H_{\alpha\beta}\partial_\gamma H_{\alpha\mu}
+2H_{\mu\alpha}\partial_\gamma H_{\alpha\beta}\partial_\mu H_{\beta\gamma}\nn \\
&\phantom{=\ }+H_{\beta\mu}\partial_\gamma H_{\alpha\beta}\partial_\alpha H_{\gamma\mu} 
-\vp \partial_\gamma H_{\alpha\beta}\partial_\alpha H_{\gamma\beta} 
-H_{\alpha\beta}\partial_\gamma H_{\alpha\beta} \partial_\mu H_{\mu\gamma}
+2H_{\alpha\beta}\partial_\gamma H_{\alpha\beta} \partial_\gamma \vp+\tfrac72b\Lambda^4\vp\,.\nn
\eeal
$\cG^0_1|_\text{old}$ coincides with the classical three-graviton vertex that one would get from expansion of the Einstein-Hilbert action \eqref{EH} using the first order expansion of the metric \eqref{gH}, except for a quantum correction,\footnote{This thus combines with the $2b\Lambda^4\vp$ to give the final term above.} $\tfrac32 b\Lambda^4\vp$,  which is generated by the action of the tadpole operator, the RHS of the linearised flow equation \eqref{flowone}, on this triple-graviton vertex  \cite{\morrii}. This quantum correction  turns $\cG_1 |_\text{old}$ into a (dimension five) eigenoperator
in standard quantisation. 

Around the Gaussian fixed point and in dimensionful variables, as in our case, an eigenoperator in standard quantisation is a local solution of the linearised flow equation \eqref{flowone}  which contains no dimensionful parameters and is polynomial in the fields. 
Since $H_{\mu\nu}c_\mu c^*_\nu$ is trivially such an eigenoperator (it has no tadpole corrections) and $\hs_0$ maps the vector space of eigenoperators into itself, as recalled below \eqref{Delta} \cite{\morrii}, we know that the $\hs_0$-exact addition \eqref{soeigenoperator} is an eigenoperator. Indeed the last term of the $\hs_0$-exact addition \eqref{soeigenoperator} is exactly right to balance the action of the (ghost) tadpole operator on $2H_{\mu\nu} c_\mu \partial_\alpha H^*_{\alpha\nu}$.
Since the relation between new and old choices \eqref{newvsold} is thus the sum of two eigenoperators of the same dimension, our new $\cG_1$ is also an eigenoperator (which of course one can also confirm by direct calculation).

%


\section{Renormalization group properties at the linearised level}
\label{sec:newquantisation}

The wrong sign $\ph$ propagator \eqref{pp} reflects the wrong sign kinetic term for $\ph$ in this gauge, which in turn is a reflection of the instability caused by the unboundedness of the Euclidean Einstein-Hilbert action (see \cite{\morri,\morrii,Morris:2018upm} for further discussion). The Euclidean partition function is then more than usually ill-defined, which the authors of ref. \cite{Gibbons:1978ac} proposed to solve by analytically continuing the $\ph$ integral along the imaginary axis. However this wrong sign does not invalidate the Wilsonian renormalization group (RG) flow equations, for example as realised by the Legendre effective action flow equation \eqref{flow}, which provide an alternative and anyway more powerful route to defining a continuum limit (see \cite{\morri,\morrii,Morris:2018upm,Kellett:2018loq} and \eg ref. \cite{Morris:1998} for further discussion).  As shown in refs. \cite{\morri,\morrii}, the wrong sign then profoundly alters the RG properties that are central to defining such a continuum limit. (For earlier observations see refs. \cite{Bonanno:2012dg,Dietz:2016gzg}.) We review and refine some of those discoveries in this section.

Consider some arbitrary infinitesimal perturbation around the Gaussian fixed point \eqref{Gzero}, whose $\ph$-amplitude dependence\footnote{\ie its $\vp$ dependence other than any dependence through space-time derivatives as in $\partial^m\!\ph$} is given by $f_\Lambda(\ph)$. 
Recalling the wrong sign in the $\vp$ propagator \eqref{pp}, and using $\dot{C}_\Lambda=-\dot{C}^\Lambda$, the linearised flow equation \eqref{flowone} implies that this \emph{coefficient function} must satisfy
\be
\label{flowf}
\dot{f}_\Lambda(\vp) = \half\,\dot{\Omega}_\Lambda\,  f^{\prime\prime}_\Lambda(\vp)\,,
\ee
where prime is $\partial_\vp$, and 
\be 
\label{Omega}
\Omega_\Lambda= |\langle \ph(x) \ph(x) \rangle | =  \int_q \frac{{C}(q^2/\Lambda^2)}{q^2} = \frac{\Lambda^2}{2a^2}
\ee
is the modulus of the $\ph$ tadpole integral regularised by the UV cutoff ($a\cu>0$ is a dimensionless non-universal constant). With the now positive sign on the right hand side of this parabolic equation, the first dramatic conclusion  
is 
that the natural direction of RG flow in this sector reverses: solutions are guaranteed to exist only when flowing from the IR towards the UV. This property will play an important r\^ole here and in later papers \cite{secondconf,second}. Most importantly, the perturbation can be written as a convergent sum over eigenoperators and their couplings only if the coefficient function is square-integrable under the corresponding Sturm-Liouville measure: 
\be 
\label{weight}
\int^\infty_{-\infty}\!\!\!\! d\ph\  {\rm e}^{\ph^2/2\Omega_\Lambda} f^2_\Lambda (\ph) <\infty\,,
\ee
where the measure is now a growing exponential. We call $\Lmm$,  the (Hilbert) space of such coefficient functions. If $f_\Lambda\in\Lmm$, then it can be written as a (typically infinite) linear combination over the operators: 
\be
\label{physical-dnL}
\dd{\Lambda}{n} := \frac{\partial^n}{\partial\vp^n}\, \dd{\Lambda}{0}\,, \qquad{\rm where}\qquad \dd{\Lambda}0 := \frac{1}{\sqrt{2\pi\Omega_\Lambda}}\,\exp\left(-\frac{\vp^2}{2\Omega_\Lambda}\right)
\ee
(integer $n\ge0$) with convergence of the sum being in the square-integrable sense under the Sturm-Liouville measure \eqref{weight}, under which also the operators are orthonormal:\notes{[856.1]}
\be 
\label{orthonormal}
\int^\infty_{-\infty}\!\!\!\! d\ph\  {\rm e}^{\ph^2/2\Omega_\Lambda}\, \dd\Lambda{n}\,\dd\Lambda{m} = \frac{n!}{\Omega_\Lambda^{\,n+1/2}\sqrt{2\pi}}\,\delta_{nm}\,.
\ee
These $\dd{\Lambda}{n}$  are solutions of the linearised flow equation for the coefficient function  \eqref{flowf}, and are nothing but the tower of non-derivative eigenoperators in the $\ph$ sector that span $\Lmm$, the general solution of the linearised flow equation in this space being a linear combination of these eigenoperators with constant coefficients, \aka couplings. The $\dd{\Lambda}{n}$
are all relevant, their scaling dimensions being equal to their engineering dimensions in mass units, namely
$-1\!-\!n$. Since $\Omega_\Lambda\propto \hbar$, the $\dd{\Lambda}{n}$ are non-perturbative in $\hbar$. It is for this reason that we must develop the theory whilst remaining non-perturbative in $\hbar$. We mention also that they are also evanescent, \ie vanish as $\Lambda\to\infty$, and have the property that the physical operators, gained by sending $\Lambda\to0$, are $\dd{}n$, the $n^\text{th}$-derivatives of the Dirac delta function.

In the $h_{\mu\nu}$ sector and the ghost sector, convergent sums are over eigenoperators that are polynomials in the fields, justifying the usual form of expansion. Altogether, the general eigenoperator can be expressed as \cite{\morrii}
\be 
\label{topFull}
\dd\Lambda{n}\, \sigma(\partial,\partial\vp,h,c,\Phi^*) +\cdots\,,
\ee
(in gauge invariant minimal basis) where we have displayed the `top term',  $\sigma$ being a $\Lambda$-independent Lorentz invariant monomial involving some or all of the components indicated, in particular the arguments $\partial\vp,h,c,\Phi^*$  can appear as they are, or differentiated any number of times. If $d_\sigma=[\sigma]$ is its engineering dimension,  then the scaling dimension of the corresponding eigenoperator is just the sum of the engineering dimensions, namely $d_\sigma\cu-1\cu-n$. Notice that undifferentiated $\vp$ does not appear in $\sigma$ but only in $\dd\Lambda{n}$. The tadpole operator in the linearised flow equation  \eqref{flowone} generates a finite number of $\Lambda$-dependent UV regulated tadpole corrections involving less fields in $\sigma$. These are the terms we indicate with the ellipses. They are formed by attaching the propagators \eqref{hh} -- \eqref{cc} (in gauge fixed basis) in all possible ways according to the usual rules of Wick contraction, but excluding $\ph$ tadpoles connected only to $\dd\Lambda{n}$, since these are already accounted for through the flow equation for the coefficient function \eqref{flowf}.

\begin{figure}[ht]
\centering
\includegraphics[scale=0.35]{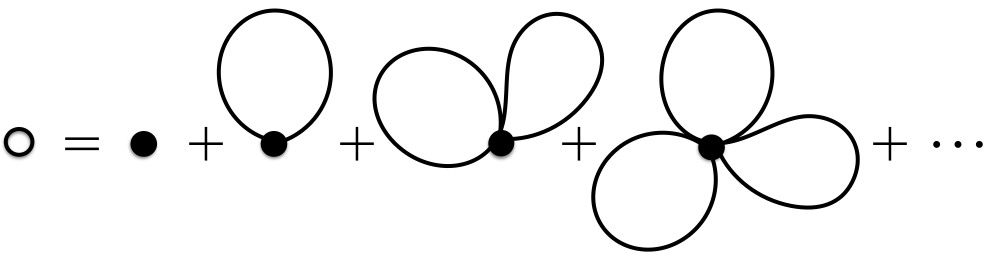}
\caption{The eigenoperator is equal to its physical limit
$\sigma\, \delta^{(n)}(\vp)$, plus all possible tadpole corrections. Those corrections generated by attaching to $\sigma$, terminate eventually (since the monomial will run out of fields), while $\vp$-tadpole corrections to $\delta^{(n)}(\vp)$ go on forever but resum to $\delta_{\, \Lambda}^{\!(n)}\!(\ph)$. We do not draw the external legs, an infinite number of which attach to $\delta_{\, \Lambda}^{\!(n)}\!(\ph)$.} 
\label{fig:tadpoles}
\end{figure}

In fact we can give the general eigenoperator \eqref{topFull} in closed form.  Note that the linearised flow equation \eqref{flowone} implies
\be 
\label{flowoneexpanded}
\dot{\Gamma}_1 = -\frac12 \dot{\prop}^{\Lambda\,AB} \frac{\partial^2_l}{\partial\Phi^B\partial\Phi^A}\,\Gamma_1\,,
\ee
where $\prop^{\Lambda\,AB} = C^\Lambda \prop^{AB}$ is the UV regulated propagator. The solution we need is therefore
\be 
\label{eigenoperatorsol}
 \exp\left(-\frac12 {\prop}^{\Lambda\,AB} \frac{\partial^2_l}{\partial\Phi^B\partial\Phi^A}\right)\, \Gamma_{1\,\text{phys}}\,, \qquad\text{where}\quad \Gamma_{1\,\text{phys}} =\sigma\, \delta^{(n)}(\vp)\,,
\ee
since at $\Lambda\cu=0$, $\dd\Lambda{n}\cu=\delta^{(n)}(\vp)$ and all the tadpole corrections vanish. The exponential operator then just generates all the Wick contractions\footnote{In particular ghost propagators count an overall $\half\times(-2)\cu=-1$ through $\langle c \bar{c}\rangle$ and $\langle \bar{c} c \rangle$ and statistics.} for the propagator which appears here as $-\prop^\Lambda$, as illustrated in fig. \ref{fig:tadpoles}. For each functional derivative we can write by the Leibniz rule
\be 
\frac{\partial_l}{\partial\Phi^A} = \frac{\partial^L_l}{\partial\Phi^A} + \frac{\partial^R_l}{\partial\Phi^A}\,
\ee
where $\partial^{L}$ acts only on the left-hand factor, here $\sigma$, and $\partial^R$ acts only the right-hand factor, here $\delta^{(n)}(\vp)$. Thus (factoring out $-C^\Lambda$ for later convenience):
\be 
\label{WickTwoFactorId}
\frac12{\prop}^{AB} \frac{\partial_l^2}{\partial\Phi^B\partial\Phi^A} = 
\frac12{\prop}^{AB} \frac{{\partial^{L}_l}^2}{\partial\Phi^B\partial\Phi^A} 
+{\prop}^{AB} \frac{\partial^L_l}{\partial\Phi^B}\frac{\partial^R_l}{\partial\Phi^A} 
+\frac12 {\prop}^{AB} \frac{{\partial^{R}_l}^2}{\partial\Phi^B\partial\Phi^A}\,.
\ee
The exponential in the eigenoperator solution \eqref{eigenoperatorsol} therefore factors into three exponentials. Since $\delta^{(n)}(\vp)$ only depends on $\vp$, the third exponential collapses to \cite{\morri}:
\be
\label{sumtadpoles}
\exp\!\left(\!-\frac12 {\prop}^{\Lambda\,AB} \frac{{\partial^{R}_l}^2}{\partial\Phi^B\partial\Phi^A}\right) \delta^{(n)}(\vp) = \mathrm{e}^{\frac12\Omega_\Lambda\partial^2_\vp}\, \delta^{(n)}(\vp) 
= \partial^n_\vp\int^\infty_{-\infty}\!\! \frac{d\vpi}{2\pi}\, \, \mathrm{e}^{-\frac12\vpi^2\Omega_\Lambda+i\vpi\vp} = \dd\Lambda{n}\,,
\ee
where we used the $\vp$ propagator \eqref{pp}, giving the tadpole integral \eqref{Omega} and derivatives $\partial_\vp$ with respect to the amplitude (\ie no longer functional), and expressed the result in conjugate momentum $\vpi$ space, after which the integral evaluates to the expression \eqref{physical-dnL} for the  $\dd\Lambda{n}$  operators. Thus the entire eigenoperator can be written as
\be 
\label{eigenoperator}
\exp\left(-{\prop}^{\Lambda\,\vp\vp} \frac{\partial^L}{\partial\vp}\frac{\partial^R}{\partial\vp}\right) \left\{ \exp\left(-\frac12 {\prop}^{\Lambda\,AB} \frac{\partial^2_l}{\partial\Phi^B\partial\Phi^A}\right)\sigma\right\} \dd\Lambda{n}\,,
\ee
where the term in braces expresses all the tadpole corrections acting purely on $\sigma$, and the left-most term generates $\vp$-propagator \eqref{pp} corrections that attach to both $\sigma$ and $\dd\Lambda{n}$ (each such attachment will increase $n\cu\mapsto n\cu+1$).

A simple example eigenoperator \cite{\morrii} will prove useful later:
\be 
\label{newtadpoleone}
-\partial_\mu c_\nu H^*_{\mu\nu}\dd\Lambda{n} +2 b\Lambda^4\dd\Lambda{n}\,.
\ee
The second term has the ghost tadpole correction  to the top monomial $\sigma\cu=-\partial_\mu c_\nu H^*_{\mu\nu}$, that we already derived in the $\hs_0$-exact addition \eqref{soeigenoperator}. (To see this immediately, substitute the $SO(4)$ decomposition \eqref{h} into the $\hs_0$-exact addition, integrate by parts, and recall the remark at the end of sec. \ref{sec:nontrivialcohomology}.) 

The continuum limit is described by the \emph{renormalized trajectory} \cite{Wilson:1973,Morris:1998}, the RG trajectory that shoots out of the (Gaussian) fixed point, parametrised by (marginally) relevant couplings that are finite at physical scales. Close to the  fixed point, the linearised approximation is justified.
The interaction there is therefore expanded only over the  marginal and relevant eigenoperators \eqref{topFull} with constant 
couplings $g^\sigma_n$ whose mass-dimensions
\be
\label{gdimension}
[g^\sigma_n] = 4-(d_\sigma\cu-1\cu-n) = 5+n-d_\sigma\,,
\ee
must all be non-negative. Every monomial $\sigma$ is therefore associated to an infinite tower of operators, which can be subsumed into
\be 
\label{firstOrder}
f^{\sigma}_\Lambda(\vp)\, \sigma(\partial,\partial\vp,h,c,\Phi^*)+\cdots = \exp\left(-{\prop}^{\Lambda\,\vp\vp} \frac{\partial^L}{\partial\vp}\frac{\partial^R}{\partial\vp}\right) \left\{ \exp\left(-\frac12 {\prop}^{\Lambda\,AB} \frac{\partial^2_l}{\partial\Phi^B\partial\Phi^A}\right)\sigma\right\} f^{\sigma}_\Lambda(\vp)\,,
\ee
where the coefficient function of the top term is given by  (at the linearised level) 
\be 
\label{coefff}
f^\sigma_\Lambda(\vp) = \sum^\infty_{n=n_\sigma} g^\sigma_n \dd\Lambda{n} \,,
\ee
and the tadpole corrections are the same as before (now with $f^\sigma_\Lambda$ differentiated according to the number of times the left-most operator acts on it).
In general all the (marginally) relevant couplings $[g^\sigma_n]\ge0$ will be needed \cite{\morri} and thus at the linearised level 
\be 
\label{nsigma} 
n_\sigma = \max(0,d_\sigma-5)\,.
\ee
For $d_\sigma\ge5$, we are thus including the marginal coupling $[g^\sigma_{n_\sigma}]=0$.




The eigenoperators (\ref{topFull},\ref{eigenoperator}) span the complete (Hilbert) space $\Ll$ of interactions whose combined amplitude dependence is square integrable under the Sturm-Liouville measure
\be 
\label{measureAll}
\exp \frac{1}{2\Omega_\Lambda}\left(\vp^2- h^2_{\mu\nu} -2\, \bar{c}_{\mu} c_{\mu}\right)\,.
\ee
At the bare level we require that $\Gamma_I$ is inside $\Ll$, so that expansion over eigenoperators is meaningful. We can interpret this as a `quantisation condition' that is thus both natural and necessary for the Wilsonian RG.  However, since we will be solving for $\Gamma_I$ directly in the continuum, our bare cutoff is already sent to infinity. Then this condition is replaced  by the requirement that  $\Gamma_I\in\Ll$ for sufficiently large $\Lambda$, where as a consequence we also have $f^\sigma_\Lambda\in\Lmm$. 

We define the \emph{amplitude suppression scale} $\Lambda_\sigma\ge0$ to be the smallest scale such that for all $\Lambda\cu>a\Lambda_\sigma$, the coefficient function is inside $\Lmm$. The coefficient function exits $\Lmm$ as $\Lambda$ falls below $a\Lambda_\sigma$, either because it develops singularities after which the flow to the IR ceases to exist, or because  it  decays too slowly at large $\vp$. 

We need to choose the $g_n^{\sigma}$ so that the flow all the way to $\Lambda\to0$ does exist, so that all modes can be integrated over and so that the physical Legendre effective action \eqref{physical} can be defined. Note that we mean by $\Gamma_\text{phys}$ the resulting $\Lambda\to0$ limit,  thus removing the infrared cutoff ($\lim_{\Lambda\to0} C_\Lambda=0$). The results are not yet physical in terms of properly incorporating diffeomorphism invariance. That requires another limit as we will shortly see.

Since the coefficient function thus exits $\Lmm$ by decaying too slowly, we know from the square-integrability condition \eqref{weight} that asymptotically: 
\be
\label{asympfLp}
f^\sigma_{a\Lambda_\sigma}(\vp)\propto A_\sigma\,\mathrm{e}^{-\vp^2/4\Omega_{a\Lambda_\sigma}+o(\vp^2)} = A_\sigma\, \mathrm{e}^{-\vp^2/2\Lambda_\sigma^2+o(\vp^2)}\,,
\ee
for at least one of $\vp\to\pm\infty$, with the other side decaying at the same rate or faster, where 
\be
\label{dimA}
[A_\sigma]=4-d_\sigma
\ee 
is a dimensionful constant, and  $o(\cdots\!)$ is a dimensionless term of either sign that grows slower than its argument. (Because of the presence of such undetermined terms, the asymptotic formula \eqref{asympfLp} only yields $A_\sigma$ up to a dimensionless proportionality constant.)  

The asymptotic behaviour \eqref{asympfLp} gives us a boundary condition which then fixes the solution of the linearised flow equation  \eqref{flowf} at large $\vp$. Thus we find (at the linearised level) the asymptotic behaviour for any $\Lambda$:
\be 
\label{largephi}
f^\sigma_\Lambda(\vp) \propto A_\sigma \exp\left(-\frac{a^2\vp^2}{\Lambda^2+a^2\Lambda^2_\sigma}+o(\vp^2)\right)
\ee
(on at least one side with the other side being the same rate or faster). From the requirement for square-integrability under the Sturm-Liouville measure, \cf \eqref{weight}, our definition of $\Lambda_\sigma$ is verified:
 $f^\sigma_\Lambda\in\Lmm$ for all $\Lambda\cu>a\Lambda_\sigma$, while $f^\sigma_\Lambda\notin\Lmm$ for $\Lambda\cu<a\Lambda_\sigma$ (in fact for all such $\Lambda$).
 
Setting $\Lambda=0$ shows that the physical coefficient function $f^\sigma_\text{phys}(\vp)$, which following \cite{\morrii} we write simply as $f^\sigma\!(\vp)$, is characterised by the decay (on at least one side with the other side being the same rate or faster):
\be 
\label{largephys}
f^\sigma\!(\vp)  \propto A_\sigma \, \mathrm{e}^{-\vp^2/\Lambda_\sigma^2+o(\vp^2)}\,.
\ee 
It appears as 
\be 
\label{fphysvertex}
f^\sigma\!(\vp)\,\sigma(\partial,\partial\vp,h,c,\Phi^*)
\ee 
in the (physical) Legendre effective  action, the regularised tadpole corrections in the $\Lambda\cu>0$ solution \eqref{firstOrder} having all vanished, since they are all proportional to positive powers of $\Lambda$. The asymptotic property for the physical coefficient function \eqref{largephys} is the motivation for calling $\Lambda_\sigma$ the amplitude suppression scale, or amplitude decay scale \cite{\morri,\morrii}.

From the linearised flow equation for the coefficient function \eqref{flowf}, this solution can be written in terms of the Fourier transform over $\vpi$:
\be 
\label{fourier-sol}
f_\Lambda^\sigma(\vp) = \int^\infty_{-\infty}\!\frac{d\vpi}{2\pi}\, \ff^\sigma\!(\vpi)\, {\rm e}^{
-\frac{\vpi^2}{2}\Omega_\Lambda+i\vpi\vp} \,, 
\ee
where $\ff^\sigma$ is $\Lambda$-independent and is thus the Fourier transform of the physical $f^\sigma(\vp)$. From the expansion of the coefficient function in terms of $\dd\Lambda{n}$ operators \eqref{coefff} and the Fourier transform expression for these operators \eqref{sumtadpoles}, the couplings are its Taylor expansion coefficients:
\be 
\label{fourier-expansion}
\ff^\sigma\!(\vpi) = \sum_{n=n_\sigma}^\infty g_n^\sigma (i\vpi)^n\,.
\ee
Since the asymptotic behaviour of the physical coefficient function \eqref{largephys} ensures that the inverse Fourier transform exists for all complex $\vpi$, $\ff^\sigma$ is an entire holomorphic function (Paley-Wiener theorem).\footnote{Then since $\ff^\sigma$ is also square integrable, the exponential decay part in the Fourier integral solution \eqref{fourier-sol} ensures that the Fourier integral converges for all complex $\vp$ provided $\Lambda>0$, and thus that $f_{\Lambda>0}^\sigma(\vp)$ is also an entire holomorphic function.} 
The asymptotic behaviour of the physical coefficient function \eqref{largephys} is reproduced by setting $\ff^\sigma\!(\vpi)$ proportional to
\be 
\label{largeFourier}
A_\sigma\Lambda_\sigma \,{\rm e}^{-\vpi^2\Lambda_\sigma^2/4+o(\vpi^2)}\,,
\ee
which also reproduces the asymptotic behaviour \eqref{largephi} at $\Lambda\cu>0$. However at this stage it needs to be interpreted with care since it captures only the fastest decaying part, corresponding to the slowest decaying behaviour in $\vp$-space. (See app. \ref{app:spectrum} for an example. This corrects part of the characterisation given in ref. \cite{\morrii}.) It does however control the large-$n$ behaviour of the couplings:
\be 
\label{largeg}
g^\sigma_n \propto A_\sigma \left(\frac{\mathrm{e}}{2n}\right)^{\frac{n}{2}}\! \Lambda^{n+1}_\sigma\,\mathrm{e}^{\,o(n)}\qquad\text{as}\quad n\to\infty\,,
\ee
where we Taylor expanded the asymptotic formula for the Fourier transform \eqref{largeFourier} and used Stirling's approximation. Indeed from the expansion of the coefficient function in terms of the $\dd\Lambda{n}$ operators \eqref{coefff}, square integrability under the Sturm-Liouville measure, as in \eqref{weight}, and the orthonormality relations \eqref{orthonormal}, we see that  
\be 
\label{converge}
\int^\infty_{-\infty}\!\!\!\! d\ph\  {\rm e}^{\ph^2/2\Omega_\Lambda} \left(f^\sigma_\Lambda\right)^2 = \frac1{\sqrt{2\pi}}\sum_{n=n_\sigma}^\infty n!\left(g^\sigma_n\right)^2 \!/\,\Omega_\Lambda^{\,n+\frac12} <\infty\qquad\text{for}\quad\Lambda>a\Lambda_\sigma\,.
\ee
By its definition, $\Lambda=a\Lambda_\sigma$ marks the radius of convergence, and thus we see that $g^\sigma_n$ must at large $n$ behave roughly like $\sqrt{\Omega^n_{a\Lambda_\sigma}/n!}$. Using Stirling's approximation we regain the asymptotic formula for the couplings \eqref{largeg} (up to sign dependence). This large-$n$ behaviour also verifies that $\ff^\sigma$ is entire.

As mentioned already below \eqref{flowf}, flows in the $\vp$-sector are guaranteed to exist in the reverse direction, \ie from the IR towards the UV. In particular, the linearised $f^\sigma_\Lambda(\vp)$ exists for all $\Lambda\ge0$ and is unique, once the coefficient function at $\Lambda=0$ is specified,  as is also clear from the Fourier integral representation \eqref{fourier-sol}. Given the asymptotic behaviour for the physical coefficient function \eqref{largephys}, this is also clear from the Green's function representation:
\be 
\label{Green}
f^\sigma_\Lambda(\vp)=\int^\infty_{-\infty}\!\!\!\! d\ph_0\,f^\sigma\!(\vp_0)\, \ddp\Lambda0{\vp\cu-\vp_0}\,. 
\ee
It is clear that this is the Green's function representation since it satisfies the linearised flow equation for the coefficient function \eqref{flowf} by virtue of the fact that the shifted eigenoperator $\ddp\Lambda0{\vp\cu-\vp_0}$ does, and returns the boundary condition in the limit $\Lambda\cu\to0$, since in this limit  $\ddp\Lambda0{\vp\cu-\vp_0}\to\delta(\vp\cu-\vp_0)$ \cite{\morri}.  Thus $\ddp\Lambda0{\vp\cu-\vp_0}$ is in fact the \emph{Heat kernel} for the diffusion equation \eqref{flowf}. By Taylor expanding $\ddp\Lambda0{\vp\cu-\vp_0}$ about $\vp$, we recover the expansion of the coefficient function over $\dd\Lambda{n}$ operators  \eqref{coefff} (and  the series converges for $\Lambda\cu>a\Lambda_\sigma$), and read off a formula for the couplings in terms of the moments of the physical coefficient function \cite{\morri}:
\be 
\label{gnfphys}
g^\sigma_n = \frac{(-)^n}{n!}\int^\infty_{-\infty}\!\!\!\!\!d\vp\,\vp^n\, f^\sigma\!(\vp)
\ee

We see therefore that the general form of the solution is given by specifying the physical coefficient function. At this stage it is subject only to the constraints that it satisfy the asymptotic condition \eqref{largephys} and be such that its Taylor expanded Fourier transform \eqref{fourier-expansion} has vanishing coefficients for $\vpi^{\,n<n_\sigma}$, equivalently that its moments \eqref{gnfphys} vanish for $n\cu<n_\sigma$.
Indeed the asymptotic property \eqref{largephys} of this $\Lambda\cu=0$ boundary condition, implies the asymptotic solution \eqref{largephi} at $\Lambda\cu>0$, which verifies that $\Lambda\cu=a\Lambda_\sigma$ marks the point above which $f^\sigma_\Lambda\in\Lmm$. Substituting the Taylor expansion formula \eqref{fourier-expansion} for the Fourier transform into the Fourier transform solution \eqref{fourier-sol} gives back the expansion of the coefficient function in terms of $\dd\Lambda{n}$ operators \eqref{coefff} which converges for $\Lambda>a\Lambda_\sigma$ and describes a valid renormalized trajectory in the linearised regime.


\section{Trivialisation in the limit of large amplitude suppression scale}
\label{sec:largeamp}

All of the above properties for the linearised solutions are inevitable consequences of respecting the wrong sign kinetic term for the conformal factor $\vp$, while insisting that the Wilsonian RG remains meaningful. 
However this general form must now be married with the first order BRST constraint \eqref{mSTone}. In ref. \cite{\morrii}, we proved that this is possible only if the coefficient function \emph{trivialises} in the sense defined below,\footnote{\label{foot:proofcorrection} In the final two paragraphs of sec. 7.2 of \cite{\morrii} we referred to ``non-constant'' coefficient functions, where we should have written ``non-trivial'' as in the current sense.} and we showed that such trivialisations are possible if
we now send $\Ls$ to infinity. In other words, we can arrange for violations of BRST to be as small as desired by taking sufficiently large $\Ls$. In this way, at first order, we get both the continuum limit and diffeomorphism invariance of the renormalized solution.



In the majority of cases the coefficient function has to become $\vp$-independent, \ie we need linearised renormalized trajectories that satisfy:
\be 
\label{flat}
f^\sigma_\Lambda(\vp) \to A_\sigma\qquad\text{as}\quad\Lambda_\sigma\to\infty\,,
\ee
(where we hold $\Lambda$, $\vp$ and $A_\sigma$ fixed and finite) such that also its $\vp$-derivatives have a limit, which is thus that they vanish.
However if BRST invariance demands a physical vertex of the same dimension but containing an undifferentiated $\vp^\alpha$ factor ($\alpha$ a positive integer), then this would appear as 
\be
\label{sigmaalpha}
\sigma =\vp^\alpha\, \sigma_\alpha(\partial,\partial\vp,h,c,\Phi^*)
\ee 
in the physical vertex \eqref{fphysvertex}, where thus the new monomial $\sigma_\alpha$ has 
\be 
\label{dsigmam}
d_{\sigma_\alpha}=d_\sigma-\alpha\,, 
\ee
and the $\vp^\alpha$ amplitude dependence must be absorbed by the physical coefficient function. 

This will correspond to linearised renormalized trajectories satisfying 
\be 
\label{flatp}
f^{\sigma_\alpha}_\Lambda(\vp) \to A_{\sigma} \left({\Lambda}/{2ia}\right)^\alpha H_\alpha\!\left({ai\vp}/{\Lambda}\right)
\qquad\text{as}\quad\Lambda_{\sigma_\alpha}\to\infty\,,
\ee
such that also their $\vp$-derivatives have a limit, where $H_\alpha$ is the $\alpha^\text{th}$ Hermite polynomial. This follows because
\be 
\label{Hermite}
\left({\Lambda}/{2ia}\right)^\alpha H_\alpha\!\left({ai\vp}/{\Lambda}\right) = \vp^\alpha + \alpha(\alpha-1)\,\Omega_\Lambda\vp^{\alpha-2}/2+\cdots
\ee
is the unique solution of the linearised flow equation for the coefficient function \eqref{flowf} with the boundary condition that it just becomes $\vp^\alpha$ at $\Lambda=0$.\footnote{These polynomials are nothing but the eigenoperators in the standard quantisation of a scalar field \cite{\morri,Bridle:2016nsu}, analytically continued along the imaginary $\vp$ axis \cite{Gibbons:1978ac}, which destroys their Hilbert space properties \cite{Dietz:2016gzg}.}

Notice that the above conditions (\ref{flatp},\ref{Hermite}) actually apply also at $\alpha=0$, where they just give back the original limit \eqref{flat} as a special case. Since we require the $\vp$-derivatives to have a limit, by l'H\^opital's rule this limit is given by the $\vp$-derivative of the right hand side. 

We say that a coefficient function \emph{trivialises} in the limit of large amplitude suppression scale if it satisfies the limiting condition \eqref{flatp} for some $\alpha$. Since at finite $\Lambda_{\sigma_\alpha}$ (with $\sigma\cu=\sigma_\alpha$), the coefficient functions satisfy the asymptotic formula \eqref{largephi}, they are \emph{non-trivial}, in particular they cannot be polynomial in $\vp$.

From the asymptotic formula for the couplings \eqref{largeg} we see that the $g^\sigma_n$ must diverge in the limit $\Lambda_\sigma\to\infty$. However the vertices are nevertheless well behaved since the coefficient function goes smoothly over to $A_\sigma$ as in the limiting condition \eqref{flat}, or more generally to the finite polynomial in \eqref{flatp}. What is happening is that the $\Lambda=a\Lambda_\sigma$ boundary, above which $f^\sigma_\Lambda(\vp)$ enters $\Lmm$, is being sent to ever higher scales. In this sense we are taking a limit towards the boundary of this Hilbert space (and thus also $\Ll$) \cite{\morrii,Morris:2018upm}. 

Actually, from the asymptotic formula for the couplings \eqref{largeg}, we can keep  $g^\sigma_n$ perturbative in this limit  if we choose $A_\sigma$ to vanish fast enough with $\Lambda_\sigma$. For example if we set $A_\sigma = a_\sigma\,\text{e}^{-\Lambda_\sigma/\mu}$ for fixed $a_\sigma$ and $\mu$, then for any finite $n$, the couplings $g^\sigma_n\to0$ as $\Lambda_\sigma\to\infty$. Although this means that the coefficient function, and thus the vertex itself, vanishes in the limit, this does not stop us from computing perturbative corrections in the usual way \cite{\morrii}, as reviewed in sec. \ref{sec:first}. We can also choose $A_\sigma$ to vanish fast enough to ensure that couplings remain uniformly perturbative (as opposed to pointwise in $n$ as in the above example). From the asymptotic formula for the couplings \eqref{largeg} one sees that for large $\Lambda_\sigma$, they first grow with $n$ and then decay once the $n^{-n/2}$ factor dominates. Thus we can estimate the maximum size coupling by differentiating with respect to $n$ and finding the stationary point. We find
\be 
\label{gmax}
g^\sigma_{n_\text{max}} \propto A_\sigma \Lambda_\sigma \,\text{e}^{\Lambda_\sigma^2/4}\qquad\text{at}\qquad n = n_\text{max} =\Lambda^2_\sigma/2\,,
\ee
which implies that we can keep the couplings uniformly perturbative if we set $A_\sigma$ to vanish faster than $\Lambda^{-1}_\sigma \,\text{e}^{-\Lambda_\sigma^2/4}$. The above result already suggests that it is the large-$n$ $g^\sigma_n$ couplings that should be important in the limit of large amplitude suppression scale. We will see this more dramatically from a different point of view in ref. \cite{second}.


\subsection{Relations} 
\label{sec:relations}

In this subsection, we pause the main development to explore two rather natural ways for generating new solutions. The first increases $\alpha$, while the second decreases it. We will see however that the maps are not inverses of each other, but rather when combined generate yet further solutions. This illustrates that there are infinitely many solutions for coefficient functions, with the same trivialisation.
The formulae we will derive are then used in the next section to arrive at the general form,  in sec. \ref{sec:examples} and app. \ref{app:spectrum} to generate examples with illustrative properties, and in sec. \ref{sec:first} to explain the properties of special limiting cases.

On the one hand, we can convert any solution to flat trivialisation limit \eqref{flat}, into one satisfying the polynomial trivialisation limit \eqref{flatp}, by multiplying the physical coefficient function by $\vp^\alpha$ and using the fact that the flow to all $\Lambda>0$ then exists and is unique. Recalling that we defined $o(\cdots\!)$ to be dimensionless, we thus identify from the asymptotic formula for the physical coefficient function \eqref{largephys}:
\be 
\label{conversion}
\Lambda_{\sigma_\alpha} = \Lambda_\sigma\qquad\text{and}\qquad A_{\sigma_\alpha} = A_\sigma \Lambda_\sigma^\alpha\,,
\ee
where $\Lambda_{\sigma_\alpha}$ is the amplitude suppression scale, and $A_{\sigma_\alpha}$ the dimensionful constant, in the asymptotic behaviour of the physical coefficient function associated to the new monomial $\sigma_\alpha$. Using the Fourier representation of the solution \eqref{fourier-sol} at $\Lambda=0$, and integration by parts, we see that the new physical coefficient function is given by setting:
\be 
\label{ffsigmap}
\ff^{\sigma_\alpha}\!(\vpi) = \left( i \partial_\vpi\right)^\alpha \ff^\sigma(\vpi)\,.
\ee
We confirm that $\ff^{\sigma_\alpha}\!(\vpi)$ thus satisfies the same general Taylor expansion formula  \eqref{fourier-expansion},  with 
\be 
\label{nsigmam}
n_{\sigma_\alpha} = \max(0,d_{\sigma_\alpha}-5)\,,
\ee
\ie defined as in the previous minimum index \eqref{nsigma},
since 
\be 
\label{nsigmap}
n_{\sigma_\alpha} = n_\sigma-\alpha = d_\sigma-\alpha-5 =  d_{\sigma_\alpha}-5\,, 
\ee
unless $d_{\sigma_\alpha}<5$ in which case $n_{\sigma_\alpha}=0$. Reading off the couplings from the Taylor expansion formula \eqref{fourier-expansion} and the Fourier transform of the new physical coefficient function \eqref{ffsigmap}, we have
\be 
\label{gsigmap}
g^{\sigma_\alpha}_{n} = (-)^\alpha (n+1)(n+2)\cdots(n+\alpha)\,  g^\sigma_{n+\alpha} = (-)^\alpha \frac{(n+\alpha)!}{n!}\, g^\sigma_{n+\alpha}\,.
\ee
Using this, the asymptotic formula for the couplings \eqref{largeg} and the conversion formulae from $\sigma$ to $\sigma_\alpha$ \eqref{conversion}, we confirm that in terms of the appropriate $\sigma_\alpha$-labelled quantities, these couplings have the expected limiting behaviour at large $n$.

On the other hand, thanks to the recurrence relation $H'_\alpha(x) =\alpha H_{\alpha-1}(x)$, one easily verifies that taking the $\vp$-derivative of the polynomial trivialisation \eqref{flatp} just maps it to ($\alpha$ times) the $(\alpha\cu-1)^\text{th}$ case, as it must since the derivative is still a solution of the flow equation for the coefficient function \eqref{flowf} and the result is determined by the physical ($\Lambda\to0$) limit, in this case $\alpha A_\sigma\vp^{\alpha-1}$. 
Of course this does not mean in general that $f^{\sigma_\alpha\,\prime}_\Lambda(\vp) = \alpha f^{\sigma_{\alpha-1}}_\Lambda(\vp)$, since there are infinitely many solutions with these limits. Indeed while $f^{\sigma_\alpha}$ 
satisfies the minimum index property \eqref{nsigmam} for each $\alpha$ in general,
the coefficient function defined by  
\be 
f^{\sigma'_{\alpha-1}}_\Lambda(\vp) := \frac1\alpha f^{\sigma_\alpha\,\prime}_\Lambda(\vp)
\ee
is more restricted. From the Fourier transform representation of the solution \eqref{fourier-sol} and its Taylor expansion \eqref{fourier-expansion} we see that it
has couplings 
\be 
\label{gsigmaprime}
g^{\sigma'_{\alpha-1}}_n = \frac1\alpha\,g^{\sigma_\alpha}_{n-1}\,, 
\ee
with the lowest $n$ in the sum thus being
\be
\label{nsigmamprime}
n_{\sigma'_{\alpha-1}} = \max(1,d_{\sigma_\alpha}-4) = \max(1,d_{\sigma_{\alpha-1}}-5)\,,
\ee
where in the last step we use the minimum index formula \eqref{dsigmam} for the $(\alpha\cu-1)^\text{th}$ case. Thus for $d_{\sigma_{\alpha-1}}\le5$, $f^{\sigma'_{\alpha-1}}$ has no $g^{\sigma'_{\alpha-1}}_0$ coupling in contrast to the general case for $f^{\sigma_{\alpha-1}}$  \viz the minimum index formula \eqref{nsigmam}.

\subsection{Simplifications and general form}
\label{sec:general}

In order to check the universal nature of the final result, we want to work with very general solutions for  linearised coefficient functions satisfying the required trivialisation constraints (\ref{flat},\ref{flatp}).
These not only determine the form of the interactions at the linearised level, but then contribute at the non-linear level through higher order contributions in the perturbative expansion \eqref{expansion}.  As will become clear \cite{secondconf}, the most powerful way to handle these higher order contributions is 
to express the solutions in conjugate momentum space. Thus we use the fact that the linearised coefficient functions are given by the Fourier transform solution \eqref{fourier-sol} via a $\Lambda$-independent $\ff^\sigma(\vpi)$ which, from its Taylor expansion \eqref{fourier-expansion} and the discussion below it, we know can be written as an entire function times a $\vpi^{\,n_\sigma}$ factor. The flat trivialisation constraint \eqref{flat} is equivalent to
\be 
\label{fflat}
\ff^\sigma(\vpi) \to 2\pi A_\sigma\, \delta(\vpi)\qquad\text{as}\quad\Lambda_\sigma\to\infty\,,
\ee
understood in the usual distributional sense (see also below) 
while more generally from the polynomial trivialisation constraint \eqref{flatp}:
\be 
\label{fflatp}
\ff^{\sigma_\alpha}(\vpi) \to 2\pi A_\sigma\, i^\alpha \delta^{(\alpha)}(\vpi)\qquad\text{as}\quad\Lambda_{\sigma}\to\infty\,,
\ee
as we see immediately from the Fourier transform flat trivialisation constraint \eqref{fflat} and the map to a Fourier transform for a coefficient function satisfying the polynomial trivialisation constraint \eqref{ffsigmap}, and which includes the flat one \eqref{fflat} as the special case $\alpha\cu=0$. (From here on for notational simplicity, we use the conversion formulae \eqref{conversion} to write $\Lambda_{\sigma_\alpha} = \Lambda_\sigma$.)

These constraints 
evidently still leave us with a huge (infinite dimensional) function space of renormalized trajectories.  We now make two further restrictions
that do not result in any significant loss of generality but greatly strengthen and streamline the analysis. 

Firstly, we insist that the coefficient functions are of definite parity, \ie even or odd functions of $\vp$. Thus those satisfying the flat trivialisation constraint \eqref{flat} will be even parity, and those satisfying the polynomial trivialisation constraint \eqref{flatp} will be even or odd, depending on whether $\alpha$ is even or odd respectively. This also implies the same of $\ff^{\sigma_\alpha}\!(\vpi)$ in the Fourier transform trivialisation constraints (\ref{fflat},\ref{fflatp}), and enforces that the asymptotic estimates for the coefficient function and its physical limit (\ref{largephi},\ref{largephys}) apply for both limits $\vp\to\pm\infty$. We see from either the expansion of the coefficient function in terms of $\dd\Lambda{n}$ operators \eqref{coefff} or the Taylor expansion of its Fourier transform \eqref{fourier-expansion}, that the couplings $g^{\sigma_\alpha}_n$ will be indexed by an integer of the same parity, and in particular the minimum index \eqref{nsigmam} required in order that the coefficient function represents a linearised renormalized trajectory, actually has this parity, so now $n_{\sigma_\alpha}$ is the smallest index of the same parity as $\alpha$ such that
\be 
\label{nsigmaalpha}
n_{\sigma_\alpha} \ge \max(0,d_{\sigma_\alpha}-5)\,.
\ee

Secondly we insist that such linearised solutions contain only one amplitude suppression scale, so that the asymptotic estimate for their Fourier transform \eqref{largeFourier} now genuinely captures their large $\vpi$ behaviour.\footnote{Examples where a spectrum of amplitude suppression scales appear were considered in ref. \cite{\morrii}, and are further developed in app. \ref{app:spectrum}.}
Then for cases satisfying flat trivialisation \eqref{fflat} we have that 
\be 
\label{ffform}
\ff^\sigma(\vpi) = 2\pi A_\sigma\,\Lambda_\sigma\, \bar{\vpi}^{n_\sigma}\, \bar{\ff}^\sigma(\bar{\vpi}^2)\,,
\ee
where $n_\sigma$ is even \ie satisfies the minimum index $n_{\sigma_\alpha}$ formula \eqref{nsigmaalpha} for $\alpha\cu=0$, $\bar{\vpi}=\Lambda_{\sigma}\vpi$ is dimensionless, and $\bar{\ff}^{\sigma}$ is a dimensionless entire function which from the asymptotic formula for the Fourier transform  \eqref{largeFourier} takes the form
\be 
\label{largevpibar}
\bar{\ff}^{\sigma}(\bar{\vpi}^2) = {\rm e}^{-\bar{\vpi}^2/4+o(\bar{\vpi}^2)}\,,
\ee
at large $\bar{\vpi}$. Likewise for general $\alpha$,
\be 
\label{ffforma}
\ff^{\sigma_\alpha}(\vpi) = 2\pi\, i^\alpha A_\sigma\,\Lambda^{\alpha+1}_{\sigma}\, \partial^\alpha_{\bar{\vpi}}\left[\bar{\vpi}^{\bar{n}_{\sigma_\alpha}}\, \bar{\ff}^{\sigma_\alpha}(\bar{\vpi}^2)\right]\,,
\ee
where $\bar{\vpi}$ has the same definition, 
and
$\bar{\ff}^{\sigma_\alpha}$ is also a dimensionless entire function satisfying the reduced asymptotic formula \eqref{largevpibar}. Note that the $\Lambda^{\alpha+1}_{\sigma}$ factor is fixed  by dimensions, \eg using the polynomial trivialisation formula  \eqref{fflatp}. Together with $A_\sigma$, these factors appear in the same form as cases satisfying flat trivialisation \eqref{ffform} if we use the identifications in the conversion formula \eqref{conversion}.

Note that the parity is carried by $\partial^\alpha_{\bar{\vpi}}$, and thus $\bar{n}_{\sigma_\alpha}$ is even. If $\alpha$ is even and $n_{\sigma_\alpha}\cu=0$ we do not require a separate $\bar{\vpi}$ power, likewise if $\alpha$ is odd and $n_{\sigma_\alpha}\cu=1$ since the $\partial_{\bar{\vpi}}$ differentials will generate a Taylor expansion with only odd powers of $\bar{\vpi}$. However if the minimum index $n_{\sigma_\alpha}$ defined in \eqref{nsigmaalpha},
is larger than these absolute minima, then the Taylor expansion of the term in square brackets must be such that all powers $\bar{\vpi}^{n>\alpha}$ are missing up to the point where we are left with an overall factor of $\bar{\vpi}^{n_{\sigma_\alpha}}$ after differentiation by $\partial^\alpha_{\bar{\vpi}}$. Without loss of generality we capture this by factoring out this power, leaving behind a function that is still entire. Thus we see that
\be 
\label{barna}
\bar{n}_{\sigma_\alpha} = 0\quad\text{if}\quad n_{\sigma_\alpha}= \varepsilon\,,
\qquad\text{otherwise}\quad
\bar{n}_{\sigma_\alpha} = n_{\sigma_\alpha}\!\!+\alpha\,,
\ee
where we define $\varepsilon=0$ or $1$ according to whether the coefficient function is even or odd.

The flat trivialisation constraint in Fourier transform space \eqref{fflatp} is then satisfied (on finite smooth functions) provided that (for $n\cu\ge0$)
\be 
\label{ffconditions}
\int^\infty_{-\infty}\!\frac{d\vpi}{2\pi}\, \frac{(i\vpi)^n}{n!} \,\ff^{\sigma_\alpha}(\vpi) \to  A_\sigma\,\delta_{n\alpha} \qquad\text{as}\quad\Lambda_\sigma\to\infty
\ee
(or we get these constraints directly from the physical limit $A_\sigma\,\vp^\alpha$, by Taylor expanding the Fourier representation \eqref{fourier-sol} in $\vp$), and from the general formula for cases satisfying the polynomial trivialisation constraint \eqref{ffforma} these are in turn satisfied if $\bar{\ff}^{\sigma_\alpha}$  is normalised as
\be 
\label{normalised}
\int^\infty_{-\infty}\!\!\!\!\!\!d\bar{\vpi}\ \bar{\vpi}^{\bar{n}_{\sigma_\alpha}}\, \bar{\ff}^{\sigma_\alpha}(\bar{\vpi}^2)\  =\ 1\,, 
\ee
and provided that for any integer $p>0$, we have
\be 
\label{vanishes}
\frac1{\Lambda^{2p}_{\sigma}}\int^\infty_{-\infty}\!\!\!\!\!\!d\bar{\vpi}\ \bar{\vpi}^{\bar{n}_{\sigma_\alpha}\!\!+2p}\, \bar{\ff}^{\sigma_\alpha}(\bar{\vpi}^2)\ \to \ 0\,, \qquad\text{as}\quad\Lambda_{\sigma}\to\infty\,.
\ee
(These integrals converge for large $\bar{\vpi}$ by virtue of the asymptotic formula  \eqref{largevpibar}.)

At first order in the perturbation theory \eqref{expansion}, $\bar{\ff}^{\sigma_\alpha}$ can be chosen to be a finite function and independent of $\Lambda_{\sigma}$, and thus the vanishing limits
 \eqref{vanishes} follow trivially. At second order in perturbation theory, we will find that we need linearised coefficient functions for which $\bar{\ff}^{\sigma_\alpha}$ depends on $\Lambda_{\sigma}$. In the majority of cases we can choose it to tend to a finite function as $\Lambda_{\sigma}\cu\to\infty$, but exceptionally it will prove useful to allow it to contain terms with coefficients that diverge logarithmically with $\Lambda_\sigma$. Clearly this mild divergence is well within the bounds implied by the vanishing limits \eqref{vanishes}.

Substituting the general formula for cases satisfying flat trivialisation \eqref{ffform} into the Fourier transform representation of the solution \eqref{fourier-sol} gives
\be 
\label{flatapproach}
f_\Lambda^\sigma(\vp) = A_\sigma \int^\infty_{-\infty}\!\!\!\!\!\!d\bar{\vpi}\ \bar{\vpi}^{\bar{n}_\sigma}\, \bar{\ff}^\sigma(\bar{\vpi}^2)\, 
\exp\left(
-\frac{\bar{\vpi}^2}{4}\frac{\Lambda^2}{a^2\Lambda_\sigma^2}+i\bar{\vpi}\,\frac{\vp}{\Lambda_\sigma}\right)\,.
\ee
Using the normalisation limit \eqref{normalised} and the vanishing limits \eqref{vanishes} we thus confirm that flat trivialisation \eqref{flat} is satisfied, and see that at large but finite $\Lambda_\sigma$ the remaining dependence is on $\Lambda^2$ and $\vp^2$ as dictated (at leading order) by dimensions and parity \viz as a Taylor series in $\Lambda^2/\Lambda^2_\sigma$ and $\vp^2/\Lambda^2_\sigma$, except for those cases at second order where such a Taylor series of corrections will also include a single factor of $\ln(\Lambda_{\sigma})$. 

Now define the polynomial function $\mathcal{H}_\alpha(\vpi,\Omega_\Lambda,\vp)$ by
\be 
(-i\partial_\vpi)^\alpha \left( {\rm e}^{
-\frac{\vpi^2}{2}\Omega_\Lambda+i\vpi\vp}\right) = \mathcal{H}_\alpha(\vpi,\Omega_\Lambda,\vp)\  {\rm e}^{
-\frac{\vpi^2}{2}\Omega_\Lambda+i\vpi\vp}\,.
\ee
Substituting the Fourier transform polynomial trivialisation constraint  \eqref{fflatp} into the Fourier transform representation of the solution \eqref{fourier-sol}, integrating by parts, and using the polynomial trivialisation definition in $\vp$-space \eqref{flatp}, we see that 
\be 
\mathcal{H}_\alpha(0,\Omega_\Lambda,\vp) = \left({\Lambda}/{2ia}\right)^\alpha H_\alpha\!\left({ai\vp}/{\Lambda}\right)\,,
\ee
where the RHS expands as given in the formula for the Hermite polynomial  \eqref{Hermite}. Thus substituting the general formula for cases satisfying the polynomial trivialisation \eqref{ffforma} into the Fourier transform representation for the solution \eqref{fourier-sol}, we have that 
\be 
\label{flatpapproach}
f_\Lambda^{\sigma_\alpha}(\vp) = A_\sigma \int^\infty_{-\infty}\!\!\!\!\!\!d\bar{\vpi}\ \bar{\vpi}^{\bar{n}_{\sigma_\alpha}}\, \bar{\ff}^{\sigma_\alpha}(\bar{\vpi}^2)\, 
 \mathcal{H}_\alpha\left(\frac{\bar{\vpi}_{\phantom{\sigma_\alpha}}}{\Lambda_{\sigma}},\Omega_\Lambda,\vp\right)\,
\exp\left(
-\frac{\bar{\vpi}^2}{4}\frac{\Lambda^2}{a^2\Lambda_{\sigma}^2}+i\bar{\vpi}\,\frac{\vp_{\phantom{\sigma_\alpha}}}{\Lambda_{\sigma}}\right)\,.
\ee
Using the normalisation limit \eqref{normalised} and the vanishing limits \eqref{vanishes} we thus confirm that polynomial trivialisation \eqref{flatp} is satisfied, and see again that the corrections are dictated by dimensions ($[\mathcal{H}_\alpha]\cu=\alpha$) and parity to be a Taylor series in $\Lambda^2/\Lambda^2_\sigma$ and $\vp^2/\Lambda^2_\sigma$, except for those cases at second order where these corrections also include 
a single factor of $\ln(\Lambda_{\sigma})$. 


We see that the difference between the left and right hand sides in polynomial trivialisation \eqref{flatp} is bounded by a term of order $1/\Lambda^2_{\sigma}$. Furthermore this is true for every relation obtained by differentiating with respect to $\vp$ on both sides until the RHS vanishes. At this point successive differentials will bring down further powers of $1/\Lambda^2_{\sigma}$ from the general finite $\Lambda_\sigma$ formula \eqref{flatpapproach} via $\vp^2/\Lambda^2_\sigma$. Thus we have for large $\Lambda_{\sigma}$: 
\beal 
\label{regularity}
\partial^p_\vp \left[ f^{\sigma_\alpha}_\Lambda(\vp) - A_{\sigma} \left({\Lambda}/{2ia}\right)^\alpha H_\alpha\!\left({ai\vp}/{\Lambda}\right) \right]\ &=\ O(1/\Lambda^2_{\sigma}) \qquad &\text{for}\quad p\le\alpha\,,\nn\\
 \partial^p_\vp f^{\sigma_\alpha}_\Lambda(\vp)\ &=\ O(1/\Lambda_{\sigma}^{2\lceil\tfrac{p-\alpha}2\rceil})\qquad &\text{for}\quad p>\alpha\,, 
\eeal
which since this applies for $p=0$, refines the earlier trivialisation characterisations (\ref{flat},\ref{flatp}), and where again one should understand that the RHS is corrected by a factor of $\ln(\Lambda_{\sigma})$ in some cases at second order. 

\subsection{Examples}
\label{sec:examples}
For example if there is no $o(\bar{\vpi}^2)$ correction in the reduced asymptotic formula \eqref{largevpibar}, then the normalisation limit \eqref{normalised} fixes the normalisation of the dimensionless entire function so that\footnote{In the case $\bar{n}_{\sigma_\alpha}=0$ one has $(-1)!!=1$.} 
\be 
\label{ffexample}
\bar{\ff}^{\sigma_\alpha}(\bar{\vpi}^2) = \frac{{\rm e}^{-\bar{\vpi}^2/4}}{(\bar{n}_{\sigma_\alpha}\!-1)!!\, 2^{\frac{\bar{n}_{\sigma_\alpha}}{2}+1} \sqrt{\pi}}\,.
\ee
In the general formula for cases satisfying flat trivialisation \eqref{ffform}, solutions to flat trivialisation \eqref{flat} that keep all possible couplings, so $n_\sigma\cu=0$, take the form
\be 
\label{ffsigmaEg}
\ff^\sigma(\vpi) = 2\pi A_\sigma\,\Lambda_\sigma\, \bar{\ff}^\sigma(\bar{\vpi}^2)\,.
\ee
Using the simplest reduced Fourier transform \eqref{ffexample} with $\alpha\cu=0$ to generate an explicit example, we have:
\be 
\label{ffoneEg}
\bar{\ff}^\sigma(\bar{\vpi}^2) =  \frac{{\rm e}^{-\bar{\vpi}^2/4}}{2\sqrt{\pi}}\,,
\ee
which just gives us our previously well-worked specimen \cite{\morri,\morrii}:
\be 
\label{foneEg}
f^\sigma_\Lambda(\vp) = \frac{a A_\sigma\Lambda_\sigma}{\sqrt{\Lambda^2+a^2\Lambda^2_\sigma}}\, {\rm e}^{-\frac{a^2\vp^2}{\Lambda^2+a^2\Lambda^2_\sigma}}\,,\quad f^{\sigma}\!(\vp) = A_\sigma\, {\rm e}^{-{\vp^2}/{\Lambda_\sigma^2}}\,, \quad g^\sigma_{2n} =\frac{\sqrt{\pi}}{n!4^n}\, A_\sigma\,\Lambda_\sigma^{2n+1}
\ee
($n=0,1,\cdots$), where the first expression follows from performing the integral in the Fourier transform representation \eqref{fourier-sol}, the second is its $\Lambda\cu\to0$ limit, and the couplings follow from the Taylor expansion relation \eqref{fourier-expansion}. 
Similarly linearised coefficient functions satisfying  $f^{\sigma_1}_\Lambda(\vp)\cu\to A_\sigma\,\vp$, with $n_{\sigma_1}=1$, have
\be 
\label{ffsigmapEg}
\ff^{\sigma_1}(\vpi) = 2\pi i A_\sigma\,\Lambda^2_{\sigma}\,\partial_{\bar{\vpi}}\, \bar{\ff}^{\sigma_1}(\bar{\vpi}^2)
\ee
from the formula for the general case \eqref{ffforma} and the reduced minimum index \eqref{barna} with $\alpha\cu=1$.
The explicit example for the simplest reduced Fourier transform \eqref{ffexample} again gives the special case \eqref{ffoneEg}, and thus
\be 
\label{fsigmaoneEg}
f^{\sigma_1}_\Lambda(\vp) = \frac{a^3\Lambda_{\sigma}^3 A_\sigma}{\left(\Lambda^2+a^2\Lambda_{\sigma}^2\right)^{3/2}}\,\vp\, {\rm e}^{-\frac{a^2\vp^2}{\vphantom{\tilde\Lambda}\Lambda^2+a^2\Lambda_{\sigma}^2}}\,,\ f^{\sigma_1}(\vp) = A_{\sigma}\,\vp \, {\rm e}^{-{\vp^2}/{\Lambda_{\sigma}^2}}\,,\ 
g^{\sigma_1}_{2n+1} = -\frac{\sqrt{\pi}}2\frac1{n!4^n}\,A_\sigma\,\Lambda_{\sigma}^{2n+3}\,,
\ee
($n=0,1,\cdots$), in agreement with coupling constant mapping formula \eqref{gsigmap} and our previously well-worked specimen \eqref{foneEg}. 
For $\alpha\cu=2$ and $n_{\sigma_2}=0$ one gets
\be 
\label{fphisquared}
f^{\sigma_2}_\Lambda(\vp) = A_\sigma \left\{ \frac{a^5\Lambda_{\sigma}^5}{\left(\Lambda^2+a^2\Lambda_{\sigma}^2\right)^{5/2}}\,\vp^2 +
\frac{a\Lambda_{\sigma}^3\Lambda^2}{2\left(\Lambda^2+a^2\Lambda_{\sigma}^2\right)^{3/2}}\right\}
{\rm e}^{-\frac{a^2\vp^2}{\vphantom{\tilde\Lambda}\Lambda^2+a^2\Lambda_{\sigma}^2}}
\ee
from the simplest reduced Fourier transform \eqref{ffexample},
which gives the physical coefficient function and couplings:
\be 
f^{\sigma_2}(\vp) = A_{\sigma}\,\vp^2 \, {\rm e}^{-{\vp^2}/{\Lambda_{\sigma}^2}}\,,\qquad
g^{\sigma_2}_{2n} = \frac{\sqrt{\pi}}2 \frac{2n\cu+1}{n!4^n}\,A_\sigma\Lambda_{\sigma}^{2n+3} \quad (n=0,1,\cdots)\,.
\ee
Its large $\Lambda_{\sigma}$ limit, $f^{\sigma_2}_\Lambda(\vp)\to A_\sigma (\vp^2+\Omega_\Lambda)$, is in agreement with polynomial trivialisation \eqref{flatp}. 

Differentiating the $\alpha\cu=1$ example \eqref{fsigmaoneEg} with respect to $\vp$:   
\be 
\label{fcheckrelation}
\check{f}^\sigma_\Lambda(\vp) = f^{\sigma_1\prime}_\Lambda(\vp)
\ee 
gives an alternative example solution for flat trivialisation \eqref{flat}: 
\be
\label{fhi2}
\check{f}^\sigma_\Lambda(\vp) = \frac{a^3\Lambda_{\sigma}^3 A_\sigma}{\left(\Lambda^2+a^2\Lambda_{\sigma}^2\right)^{3/2}}\left(1-\frac{2a^2\vp^2}{\vphantom{\tilde\Lambda}\Lambda^2+a^2\Lambda_{\sigma}^2}\right) {\rm e}^{-\frac{a^2\vp^2}{\vphantom{\tilde\Lambda}\Lambda^2+a^2\Lambda_{\sigma}^2}}\,,\quad 
\check{f}^{\sigma}\!(\vp) = A_\sigma\, \left(1-\frac{2\vp^2}{\Lambda^2_\sigma}\right) {\rm e}^{-{\vp^2}/{\Lambda_\sigma^2}}
\ee
as is clear from the large amplitude suppression scale limit. However this solution has $\check{g}^\sigma_0\cu=0$ as is immediately clear from integrating the $\check{f}$ relation \eqref{fcheckrelation} and using the moment relation \eqref{gnfphys}.  In sec. \ref{sec:relations} we showed that $n_\sigma\cu=1$  -- or rather $n_\sigma\cu=2$ since it is even, \cf the general $n_{\sigma_\alpha}$ definition \eqref{nsigmaalpha}. Indeed differentiating the Fourier transform representation \eqref{fourier-sol}, and using the general $\alpha\cu=1$ Fourier transform \eqref{ffsigmapEg} and the simplest normalised reduced form \eqref{ffoneEg}, we see that the corresponding $\check{\ff}^\sigma(\vpi)$ takes the general form for cases satisfying flat trivialisation \eqref{ffform}:
\be 
\label{ffcheckEg}
\check{\ff}^\sigma(\vpi) = 2\pi A_\sigma\,\Lambda_\sigma\, \bar{\vpi}^2\,\check{\bar{\ff}}^\sigma(\bar{\vpi}^2)\,,\qquad
\check{g}^\sigma_{2n} =-\frac{2\sqrt{\pi}}{(n\cu-1)!\,4^n}\, A_\sigma\,\Lambda_\sigma^{2n+1}\,, 
\ee
if $\check{\bar{\ff}}^\sigma = \bar{\ff}^\sigma/2$, \cf the $\alpha\cu=1$ example \eqref{ffoneEg}, in agreement with the simplest reduced Fourier transform  \eqref{ffexample}. Expanding in $\vpi$ and using the Taylor expansion formula \eqref{fourier-expansion} then yields the displayed couplings, in agreement with the coupling constant mapping formula \eqref{gsigmap} (and actually the above formula holds also for $n\cu=0$ if we interpret $(-1)!$ as the Euler $\Gamma(0)=\infty$). 
Finally, notice that in all these examples, the approach to the trivialisation limits (\ref{flat},\ref{flatp}) is as described at the end of sec. \ref{sec:general}.

\section{Continuum limit at first order in perturbation theory}
\label{sec:first}

We will treat the first order cosmological constant term, associated to its BRST cohomology representative \eqref{Gonecc}, at the end of this section.
The remaining parts of $\cG_1$ that we computed in \eqref{Gonetwo}, \eqref{Goneone} and \eqref{Gonezero} will 
provide us with the top monomials $\sigma$ that we need to construct the derivative part. In order to be supported on the renormalized trajectory, such that $\Gamma_1$ is constructed, these $\sigma$ need to be `dressed' with coefficient functions $f^\sigma_\Lambda(\vp)$ as in the general closed formula for the eigenoperator \eqref{firstOrder}.
In the most general case we should give each top term its own coefficient function. This would provide the most complete test of universality of the continuum limit, however at the expense of carrying around a lot more terms and labels. At sufficiently high order of perturbation theory in the perturbative expansion \eqref{expansion}, we expect to have to do this because these $\Gamma_1$ couplings will then run independently \cite{\morrii}. In fact we will show in ref. \cite{secondconf} that as a consequence of specialising to coefficient functions of definite parity, the $\Gamma_1$ couplings do not run at second order but  they can be expected to run at third order. 

Here it is not necessary to treat the general case, since we will see that the passage to universality is very generic such that 
it is clear that this will continue to work when we give each top monomial in $\Gamma_1$ its own coefficient function.
We thus find that for our purposes just two coefficient functions are sufficient for constructing $\Gamma_1$, the first of which we label as $f^1_\Lambda(\vp)$, setting the superscript to $\sigma=1$ \ie the perturbation level index, and in the second case choose the label $\sigma=1_1$ as in $\alpha\cu=1$ trivialisation \eqref{flatp} to indicate that $f^{1_1}_\Lambda(\vp)$ absorbs a factor of $\vp$. Thus $f^1_\Lambda(\vp)$ is even, while $f^{1_1}_\Lambda(\vp)$ is odd.  
Although in principle every vertex can have its own amplitude suppression scale $\Lambda_\sigma$, we will find that we can choose them all to be equal. To make clear that it is independent of $\sigma$, we set this common amplitude suppression scale to $\Lambda_\sigma=\Lambda_\p$ (borrowing the notation already used in \cite{\morri}). 


Now since $\cG_1$ is a dimension $d_1=5$ operator, we have by dimensions  \eqref{dimA} that the dimensionful coefficient $[A_{1}]=-1$. As the remaining factor in front $\cG_1$, after taking the limit $\Lambda_\p\to\infty$, we recognise that it is actually $A_1=\kappa$, where the latter was defined in \eqref{kappa},
\ie we have
\be 
\label{fonelimit}
f^1_\Lambda(\vp) \to \kappa\,,\qquad f^{1_1}_\Lambda(\vp)\to\kappa\,\vp\,,
\qquad\text{as}\quad\Lambda_\p\to\infty\,,
\ee
where whenever we now write the limit of large amplitude suppression scale, we mean also the more refined regularity properties \eqref{regularity}, in particular in these cases the limits are reached at least as fast as $1/\Lambda_\p^2$.
We see that Newton's constant therefore arises only as a kind of collective effect of all the renormalizable couplings $\{g^1_{2n},g^{1_1}_{2n+1}\}$, these latter being responsible for forming the continuum limit. Indeed $A_1\cu=\kappa$ is not an underlying coupling in its own right but rather appears as the overall proportionality constant when the couplings are expressed in terms of $\Lambda_\p$, through their asymptotic formula \eqref{largeg}. 

Examples of such coefficient functions were given in \cite{\morrii} and appear in equations \eqref{foneEg} and \eqref{fsigmaoneEg}. We stress however that we are working here with very general solutions for these coefficient functions. From the definition of the minimum index $n_{\sigma_\alpha}$ \eqref{nsigmaalpha} and the expansion of the coefficient function over the operators $\dd\Lambda{n}$ \eqref{coefff}, we have that in general all eigenoperators will be involved: 
\be 
\label{coeffone}
f^1_\Lambda(\vp) = \sum^\infty_{n=0} \,g^1_{2n}\, \dd\Lambda{2n}\,, \qquad f^{1_1}_\Lambda(\vp) = \sum^\infty_{n=0}\, g^{1_1}_{2n+1}\, \dd\Lambda{2n+1}\,,
\ee
where these sums converge (in the square integrable sense) for $\Lambda>a\Lambda_\p$. From the general dimension formulae \eqref{gdimension}, and \eqref{dsigmam}:
\be 
\label{gonedimensions}
[g^1_{2n}] = 2n\,,\qquad [g^{1_1}_{2n+1}] = 2n+2\,.
\ee
Thus all these couplings are relevant, with the exception of $g^1_0$ which is marginal. Up to second order it does not run \cite{secondconf} and thus behaves as though it is exactly marginal, parametrising a line of fixed points. 



From the antighost level two free BRST cohomology representative \eqref{Gonetwo}, we thus set at antighost level two:
\be 
\label{Gammaonetwo}
\Gamma^2_1 = -c_\nu\,\partial_\nu c_\mu\, c^*_\mu\, f^1_\Lambda(\vp)\,.
\ee
Since $f^1_\Lambda$ is taken to satisfy the linearised flow equation for coefficient functions \eqref{flowf} and there is no other opportunity to attach tadpoles to \eqref{Gammaonetwo}, $\Gamma^2_1$ already satisfies the linearised flow equation \eqref{flowone}, and thus appears correctly as a sum over eigenoperators. Evidently at this antighost level, the linearised mST \eqref{mSTone} is satisfied in the limit (by the more refined limits \eqref{regularity} at least as fast as $1/\Lambda^2_\p$) since:
\be
 Q_0\,\Gamma^2_1  = -c_\nu\,\partial_\nu c_\mu\, c^*_\mu\, \partial\cu\cdot c\,f^{1\prime}_\Lambda(\vp)  \to0 \qquad\text{as}\quad\Lambda_\p\to\infty\,,
 \ee 
and $\Gamma^2_1 \to \kappa\,\cG^2_1$ then coincides with a legitimate choice in the usual perturbative quantisation. 

As discussed above sec. \ref{sec:relations}, if we keep $\kappa$ fixed in the large amplitude suppression scale limit, all the couplings $\{g^1_{2n},g^{1_1}_{2n+1}\}$ diverge. As we noted however,  we can stay perturbative by requiring instead that $\kappa$ vanish fast enough. Although this makes the vertex vanish, we can still extract the same results by phrasing  the limit more carefully as $\Gamma^2_1/\kappa\to\cG^2_1$.  From here on we will take this phrasing as tacitly understood.\footnote{This is in conformity with the reasonable assumption that the expansion in $\kappa$ is only asymptotic \cite{\morrii}. Then strictly speaking the expansion only anyway makes sense in the $\kappa\to0$ limit, \ie as Taylor expansion coefficients in $\kappa$ .} 

In the antighost level one free BRST cohomology representative \eqref{Goneone} we need to substitute the $SO(4)$ decomposition \eqref{h} into the last term to isolate the factor of $\vp$, and thus the dressed antighost-level-one piece appears as
\be 
\label{Gammaoneone}
\Gamma^1_1 = - \left(c_\alpha \partial_\alpha H_{\mu\nu} + 2\, \partial_\mu c_\alpha h_{\alpha\nu}\right) H^*_{\mu\nu}\, f^1_\Lambda(\vp) -\partial_\mu c_\nu H^*_{\mu\nu}\, f^{1_1}_\Lambda (\vp)\,.
\ee
This time the result does not yet satisfy the linearised flow equation \eqref{flowone}, 
unlike with the previous choice in ref. \cite{\morrii}, because it requires the tadpole correction in the $\hs_0$-exact eigenoperator \eqref{soeigenoperator} or rather as formulated for the new quantisation in \eqref{newtadpoleone}.\footnote{Tadpole contributions from the first term in the dressed antighost-level-one piece \eqref{Gammaoneone} all vanish, either because the tadpole integral is odd in momentum or because $h_{\alpha\alpha}=0$.} In other words the sum over eigenoperators is actually $\Gamma^1_1+2b\Lambda^4 f^{1_1}_\Lambda(\vp)$.
Since $\Delta^-\,\Gamma^2_1$ trivially vanishes, the descendant equation \eqref{descendants} that relates $\Gamma^2_1$ to $\Gamma^1_1$ reads:
\besp 
\label{descendantonetwo}
Q_0^-\,\Gamma^2_1+Q_0\,\Gamma^1_1 = 
-\partial_\mu c_\nu \, \partial\cu\cdot c\, H^*_{\mu\nu}\left(f^1_\Lambda-f^{1_1\prime}_\Lambda\right)\\
- 2( c_\alpha\partial_\alpha c_\mu)H^*_{\mu\nu}\partial_\nu\vp f^{1\prime}_\Lambda  
- \left(c_\alpha \partial_\alpha H_{\mu\nu} + 2\, \partial_\mu c_\alpha h_{\alpha\nu}\right) H^*_{\mu\nu}\,  \partial\cu\cdot c \,f^{1\prime}_\Lambda\,,
\eesp
where we used the Koszul-Tate charge \eqref{KTHc} and note from the free BRST transformation \eqref{QH} that 
\be 
Q_0\, h_{\mu\nu} = \partial_\mu c_\nu +\partial_\nu c_\mu -\half \,\delta_{\mu\nu}\, \partial\cu\cdot c\,.
\ee
It is clear from the first-order coefficient function trivialisation formulae \eqref{fonelimit}  that as required $Q_0^-\,\Gamma^2_1+Q_0\,\Gamma^1_1\to0$ (at least as fast as $1/\Lambda^2_\p$). At the expense of some generality, we could eliminate the first term on the RHS of the descendant equation \eqref{descendantonetwo} by setting 
\be 
\label{simplifyf}
f^1_\Lambda = f^{1_1\prime}_\Lambda\,. 
\ee 
By the minimum index map \eqref{nsigmamprime} this would also eliminate $\dd\Lambda0$, \ie set $g^1_0=0$. We would still be left with the $\Lambda_\p\cu<\infty$ violations on the second line however.

Finally, extracting the undifferentiated $\vp$ pieces from the antighost level zero free BRST cohomology representative \eqref{Gonezero} by using the $SO(4)$ decomposition \eqref{h}, we have as in \cite{\morrii} that the first order graviton interaction is made up of twelve top terms and one tadpole contribution:
\besp 
\label{Gammaonezero}
\Gamma^0_1 = \Big( 
\frac14h_{\alpha\beta}\partial_\alpha\vp\partial_\beta\vp
-h_{\alpha\beta}\partial_\gamma h_{\gamma\alpha}\partial_\beta\vp
-\frac12 h_{\gamma\delta}\partial_\gamma h_{\alpha\beta}\partial_\delta h_{\alpha\beta}
-h_{\beta\mu}\partial_\gamma h_{\alpha\beta}\partial_\gamma h_{\alpha\mu} \\
+2h_{\mu\alpha}\partial_\gamma h_{\alpha\beta}\partial_\mu h_{\beta\gamma} 
+h_{\beta\mu}\partial_\gamma h_{\alpha\beta}\partial_\alpha h_{\gamma\mu}
-h_{\alpha\beta}\partial_\gamma h_{\alpha\beta} \partial_\mu h_{\mu\gamma}
+\frac12h_{\alpha\beta}\partial_\gamma h_{\alpha\beta} \partial_\gamma \vp \Big) f^1_\Lambda \\
+\left( \frac38(\partial_\alpha\vp)^2
-\frac12\partial_\beta h_{\beta\alpha}\partial_\alpha\vp
-\frac14(\partial_\gamma h_{\alpha\beta})^2 
+\frac12 \partial_\gamma h_{\alpha\beta}\partial_\alpha h_{\gamma\beta} \right) f^{1_1}_\Lambda
+\frac72 b \Lambda^4 f^{1_1}_\Lambda \,,
\eesp
except that the tadpole contribution now appears with coefficient $\frac72 =2\cu+\frac32$. 
The final descendant equation \eqref{descendants} is satisfied in the limit:
\be 
\label{descendantoneone}
Q_0\,\Gamma^0_1 +\left(Q^-_0-\Delta^-\right)\Gamma^1_1-\Delta^=\,\Gamma^2_1\to 0\,,
\ee
at least as fast as $1/\Lambda^2_\p$, since the individual limits are also reached at least as fast as $1/\Lambda^2_\p$\,:
\be 
\label{Gammaonelimits}
\Gamma^n_1\to \kappa\,\cG^n_1\,,\qquad \text{as}\quad \Lambda_\p\to\infty\,.
\ee
It is straightforward to verify the above descendant equation \eqref{descendantoneone} directly. 
To evaluate \eg $\Delta^-\,\Gamma^1_1$, one inverts the $SO(4)$ decomposition \eqref{h} to give $h_{\mu\nu}=H_{\mu\nu}-\frac14\delta_{\mu\nu}H_{\alpha\alpha}$ and $\vp=\half H_{\mu\mu}$, or recognises that \cite{\morrii}
\be 
\frac{\partial}{\partial H_{\alpha\beta}} = \frac{\partial h_{\mu\nu}}{\partial H_{\alpha\beta}}\frac{\partial}{\partial h_{\mu\nu}} + \frac{\partial\vp}{\partial H_{\alpha\beta}} \frac{\partial}{\partial\vp}
= \frac{\partial}{\partial h_{\alpha\beta}} + \frac12\,\delta_{\alpha\beta}\, \frac{\partial}{\partial\vp}\,.
\ee
Note that although these measure terms give contributions proportional to some positive power of $\Lambda$, thanks to UV regularisation by $C$, for example 
\be 
\label{DmmGammaonetwo}
-\Delta^=\,\Gamma^2_1 = -b\Lambda^4\partial\cu\cdot c f^1_\Lambda\,,
\ee
it does not alter the speed at which they vanish in the limit of large $\Lambda_\p$ (as can be verified here by integration by parts).

In the opposing limits there is no sense in which a non-trivial diffeomorphism invariance holds because the dependence on the conformal factor forbids it \cite{\morrii}. For example if $\vp\gg \Lambda_\p,\Lambda$, the coefficient functions are no longer given approximately by $\kappa$ and $\kappa\vp$, but rather take the exponentially decaying form demanded by the asymptotic formula \eqref{largephi}.

These statements hold also if we express everything in dimensionless variables using $\Lambda$, as needed to clearly see the Wilsonian RG behaviour \cite{\morri,Morris:1998}. We write dimensionless variables with a tilde, so \eg $\tilde{q}^\mu = q^\mu/\Lambda$, $\tp=\vp/\Lambda$, whilst we write $\delta_{n}(\tp) = \Lambda^{1+n} \,\dd\Lambda{n}$ for the scaled operator \cite{\morri}. The dimensionless couplings run with $\Lambda$ according to their mass dimensions \eqref{gonedimensions}:
\be 
\label{dimensionless}
\tg^1_{2n}(\Lambda) = g^1_{2n}/ \Lambda^{2n}\,, \qquad \tg^{1_1}_{2n+1}(\Lambda) = g^{1_1}_{2n+1}/\Lambda^{2n+2}\,.
\ee
We thus confirm that $\Gamma$ approaches the Gaussian fixed point ($g^1_0=0$) or more generally the line of fixed points $g^1_0=\tg^1_0\ne0$, as $\Lambda\to\infty$. In particular all the relevant parts of $\Gamma_1$ vanish as negative powers of $\Lambda$, with non-trivial $\tp$ dependent coefficients being the corresponding scaled operator $\delta_{2n+\eps}(\tp)$. In the limit only the marginal contribution $\tilde{f}^1_\Lambda(\tp)\to g^1_0\, \delta_0(\tp)$ in this sole coefficient function survives (and still carries non-trivial $\tp$ dependence). 

In dimensionful variables, 
if $\Lambda$ is much larger than the other scales $\Lambda_\p,\vp$, the situation is a little obscured but it is still the case that there is no sense in which a non-trivial diffeomorphism invariance is recovered. 
The coefficient functions are again dominated by the lowest terms in the expansion \eqref{coeffone}. Using the explicit formulae for the $\dd\Lambda{n}$ operators \eqref{physical-dnL} we have
in the current case \notes{Maple/QG/evanescents2}
\beal 
f^1_\Lambda &= \frac{a }{\Lambda\sqrt{\pi}}\,g^1_0 -\frac{a^3}{\Lambda^3\sqrt{\pi}}\left(g^1_0\vp^2+2g^1_2\right) + \frac{a^5}{2\Lambda^5\sqrt{\pi}}\left(g^1_0\vp^4+12g^1_2\vp^2+24g^1_4\right) 
+O\left(\frac1{\Lambda^7}\right)\nn\\ 
f^{1_1}_\Lambda &= -\frac{2a^3 }{\Lambda^3\sqrt{\pi}}\,g^{1_1}_1\,\vp +
\frac{2a^5}{\Lambda^5\sqrt{\pi}}\left(g^{1_1}_1\vp^3+6g^{1_1}_3\right) 
+O\left(\frac1{\Lambda^7}\right)\,.
\label{largeLambdaf}
\eeal 
The leading terms, and only the leading terms, have the correct $\vp$ dependence to allow BRST invariance to be recovered, however with $g^1_0\ne0$ they have the wrong ratio. (They should have equal coefficients, but this is impossible at diverging $\Lambda$ since $g^1_0$ and $g^{1_1}_1$ must be fixed and finite.) By setting $g^1_2=g^{1_1}_1$, and $g^1_0=0$, (only) the leading terms have both the correct $\vp$ dependence and the  correct ratio, as in fact would result from the identification \eqref{simplifyf} of the two coefficient functions, \cf the coupling constant mapping formula \eqref{gsigmaprime}, although with an effective $\kappa$ that then vanishes as $\kappa_\text{eff}\sim 1/\Lambda^3$. Meanwhile the measure terms in the above descendant formula \eqref{descendantoneone}  provide divergent obstructions to satisfying $\hs_0\,\Gamma_1=0$, if $g^1_0\ne0$. Thus evaluating the measure term formula \eqref{DmmGammaonetwo} tells us that
\be 
-\Delta^=\,\Gamma^2_1  = \Lambda \frac{ba^3}{\sqrt{\pi}}\,g^1_0\,\partial\cu\cdot c\,\vp^2 +O\left(\frac1{\Lambda}\right)\,,
\ee
(dropping total derivative terms), and $\Delta^-\Gamma^1_1$ provides also such a term but with coefficient $-\frac92$ and also a $g^1_0\Lambda(c_\alpha \partial_\alpha\vp+\partial_\alpha c_\beta h_{\alpha\beta})\vp$ piece arising from the contribution containing $\Delta^-(H^*_{\mu\nu} f^1_\Lambda)$. Setting $g^1_0=0$ removes these divergences but leaves us with subleading terms that violate BRST invariance, as is also true of the subleading terms in the large $\Lambda$ expansion of the coefficient functions
 \eqref{largeLambdaf} in this case.

This completes the demonstration at first order. The result fits the picture we sketched in the Introduction, \cf fig. \ref{fig:flow}. In particular for $\Lambda_\p\gg \Lambda,\vp$, diffeomorphism invariance holds in the sense that 
\be
\hs_0 \Gamma_1 = \hs_0\, (\Gamma^2_1+\Gamma^1_1+\Gamma^0_1) = O(1/\Lambda^2_\p)\,.
\ee
This means in particular in the limit $\Lambda_\p\to\infty$ and the physical limit ($\Lambda\to0$), we recover diffeomorphism invariance precisely in terms of satisfying the standard Slavnov-Taylor (Zinn-Justin) identities, namely at first order $(Q_0+Q_0^-)\Gamma_1 = (\Gamma_0,\Gamma_1) = 0$, where we used the general definition of the charges \eqref{charges}, the linearised mST \eqref{mSTone} and noted that from the definition of the measure operator \eqref{Delta} that $\Delta\to0$ as $\Lambda\to0$.

Finally, we remark that including a cosmological constant is straightforward at first order. We need to dress its BRST cohomology representative \eqref{Gonecc} with its own coefficient function. Since we must absorb the factor of $\vp$, the monomial $\sigma\cu=1$ is simply the unit operator, whilst we must choose an odd coefficient function $f^{cc}_\Lambda(\vp)$ with the trivialisation
\be 
f^{cc}_\Lambda(\vp)\to \lambda\,,\qquad\text{as}\quad\Lambda_\p\to\infty\,,
\ee
where $\kappa^2\lambda/4$ is the standard cosmological constant.  At this order we do not need a whole separate odd coefficient function and can by the trivialisation property \eqref{fonelimit} for $f^{1_1}$, just set $f^{cc}_\Lambda = \lambda f^{1_1}_\Lambda/\kappa$. The linearised mST \eqref{mSTone} is satisfied in the limit because $Q_0 f^{cc}_\Lambda(\vp) = \partial\cu\cdot c\, f^{cc\,\prime}_\Lambda(\vp)\to0$ at least as fast as $1/\Lambda_\p^2$, as follows by integration by parts and using the refined limits \eqref{regularity}, or directly by the observation that the first order vertex tends to $\kappa$ times its free BRST cohomology representatives, \viz \eqref{Gammaonelimits}. Indeed these properties were already used in proving the invariance (in the limit) of the last term in the antighost level zero part of the first order vertex \eqref{Gammaonezero}.

\section{Discussion}
\label{sec:discussion}

In this section we discuss further the meaning and implications of this construction and draw out its relation to other approaches. As recalled at the beginning of sec. \ref{sec:newquantisation}, the Euclidean signature Einstein-Hilbert action is unbounded below. From sign of the action \eqref{EH}, the instability is towards manifolds of arbitrarily positive curvature. Whilst this conformal factor instability \cite{Gibbons:1978ac} means that the partition function is not well defined, the Wilsonian exact RG flow equation remains well defined \cite{Reuter:1996,\morri}, and anyway provides a more powerful route towards constructing the continuum limit. However the wrong sign propagator \eqref{pp} for the conformal factor ($\vp$), has a profound effect on RG properties. Close to the UV Gaussian fixed point, \cf fig. \ref{fig:flow}, the  requirement that expansion over eigenoperators converges, picks out the Hilbert space $\Lmm$ defined by the Sturm-Liouville measure \eqref{weight}, 
which is spanned by the novel set of eigenoperators \eqref{physical-dnL}, the $\dd\Lambda{n}$. 

We must emphasise that the requirement that one works within $\Lmm$ (more generally $\Ll$ defined by \eqref{measureAll}, when the other fields are included) is crucial for the Wilsonian RG to make sense in an otherwise unrestricted space of functions (of $\vp$). Without this restriction the eigenoperator spectrum degenerates, becoming continuous, and it is no longer possible to unambiguously divide a perturbation into its relevant and irrelevant parts \cite{Dietz:2016gzg}. This problem lay unnoticed until ref. \cite{Dietz:2016gzg} and as yet has only been further addressed in refs. \cite{Gonzalez_Martin_2017,\morri,Kellett:2018loq,Morris:2018upm,\morrii}. The reason that it lay undiscovered is primarily because to see this problem of convergence one must work with solutions involving an infinite number of operators (the exact solution being also of this type). However, with few prior exceptions \cite{Reuter:2008wj,Bonanno:2012dg,Dietz:2012ic}, quantum gravity investigations using exact RG flow equations worked within truncations (model ans\"atze) where only a finite number of operators are retained. 

Restricting flows to the diffeomorphism invariant subspace, \cf fig. \ref{fig:flow}, might be expected to solve the problem since diffeomorphism invariance at the classical level restricts the functional dependence on the conformal factor to just a few operators at any given order in the derivative expansion. However when carefully analysed, the so-called $f(R)$ approximations  \cite{Machado:2007,Codello:2008,Benedetti:2012,Demmel:2014hla,Demmel2015b,Ohta:2015efa,Ohta2016,Percacci:2016arh,Morris:2016spn,Falls:2016msz,Ohta:2017dsq},
which are diffeomorphism invariant model ans\"atze that keep an infinite number of operators, also show the problem that the eigenoperator spectrum degenerates \cite{Dietz:2012ic,Gonzalez_Martin_2017}, and furthermore it is now clear that the underlying cause is the conformal factor instability \cite{Dietz:2012ic,Dietz:2016gzg}. Indeed it was these problems that motivated the studies \cite{Dietz:2015owa,Dietz:2016gzg}.

Within standard perturbation theory the problem can be ignored, the wrong sign $\vp$ propagator \eqref{pp} being apparently harmless. As recalled in sec. \ref{sec:newquantisation}, the conformal factor instability was identified in ref. \cite{Gibbons:1978ac}, where they proposed to solve it by analytically continuing the $\ph$ integral along the imaginary axis. This does not alter final perturbative results, but non-perturbatively it is less clear that this treatment makes sense \cite{Feldbrugge:2017mbc}.
Some other approaches keep, and seek to cope with, the conformal factor instability (but do not treat the convergence problems whose solution leads uniquely to our proposal).  In ref. \cite{Bonanno:2013dja} a model truncation to a finite set of operators, ``$-R\cu+R^2$ '' gravity, was considered within the non-perturbative asymptotic safety scenario \cite{Reuter:1996}.  
The right-sign $R^2$ term stabilises the conformal sector, resulting in an unsuppressed non-perturbative Planckian scale modulated phase which breaks Lorentz symmetry. If physical, this would be  phenomenologically challenging \cite{Pospelov_2004,Pruttivarasin_2015,Evans:2015zwa}.
A somewhat similar effect is seen in the Causal Dynamical Triangulations approach to quantum gravity \cite{Loll:2019rdj}. Although a restriction here to a global time foliation leads to an encouraging phase structure, the conformal instability towards a crumpled phase remains, and this programme has yet to succeed in furnishing an acceptable continuum limit \cite{Ambjorn:2020rcn}.


Returning to our paper, the fact that $[\dd\Lambda{n}]=-1\cu-n$ form a tower of increasingly relevant operators, implies that all interactions  are dressed with coefficient functions $f^\sigma_\Lambda(\vp)$ which contain an infinite number of  relevant underlying couplings, $g^\sigma_n$. Close to the Gaussian fixed point, the linearised flow equation \eqref{flowf} is justified. Then, as we showed in sec. \ref{sec:newquantisation}, and also in \cite{\morri}, if $f^\sigma_\Lambda(\vp)\in\Lmm$, it is guaranteed to remain there at all higher scales. Thus the requirement that for sufficiently high $\Lambda$ we have $f^\sigma_\Lambda(\vp)\in\Lmm$, can be seen as a quantisation condition that is both natural and necessary for the Wilsonian RG.

Note that in this step we are relying on the fact that the Cauchy initial value problem itself is well defined in the UV direction \cite{Bonanno:2012dg,Dietz:2016gzg,\morri}, \ie the property that the RG flow is guaranteed to exist to all higher scales. This is the reverse direction from normal: another consequence of the wrong sign $\vp$-propagator.  However the fact that the well defined flow direction is now opposite to the one defined by integrating out microscopic degrees of freedom, is an example where, even for the Wilsonian RG equation, the wrong sign $\vp$-propagator forces us to reassess some of the usual physical intuition.
We emphasise that this property does not alter the fact that the bare action determines, eventually after integration over all momentum modes, and up to universality, the physical effective action \eqref{physical}. Rather it throws obstacles in the path towards constructing this, that have not been previously encountered or recognised as such. Thus for example for a generic choice of bare coefficient function $f^\sigma_{\Lambda_0}(\vp)$ at an initial UV scale $\Lambda\cu=\Lambda_0$, the flow to the IR will almost certainly fail at some finite critical scale $0\cu<\Lambda\cu=\Lambda_{cr}\cu<\Lambda_0$ after which it ceases to exist \cite{\morri}. Since one is then unable to complete the integration over all modes, the quantum field theory as a physical entity  itself ceases to exist in this case \cite{\morri}. 

As we saw the coefficient functions that do survive all the way to the IR have a physical limit \eqref{largephys} which decays for large $\vp$ with some characteristic amplitude suppression scale, $\Lambda_\sigma$. Even for such coefficient functions, if $\Lambda_\sigma$ is finite, the complete flow and thus also the physical theory, can cease to exist on sufficiently small and asymmetrical manifolds \cite{\morri,Kellett:2018loq}. Tantalising as this seems \cite{\morri,Kellett:2018loq}, in order to recover diffeomorphism invariance we need the coefficient functions to trivialise, \cf sec. \ref{sec:largeamp}, and in practice this requires taking the limit $\Lambda_\sigma\cu\to\infty$ in the continuum theory \cite{\morrii} (holding everything else fixed). Then the above restrictions on the allowed manifold  \cite{\morri,Kellett:2018loq} appear to be ruled out except possibly to rule out manifolds with singularities \cite{\morrii}. The amplitude suppression scale per se should therefore be seen as part of the procedure for forming the continuum limit and not as having direct influence on the physical theory. Nevertheless it is the cross-over scale that matches the RG flow in the diffeomorphism invariant subspace to the upper part of the renormalized trajectory, \cf fig. \ref{fig:flow}, and as such plays a r\^ole in determining which of these RG flows actually correspond to a valid perturbative continuum limit. It may also leave behind certain finite logarithmic corrections at higher order in perturbation theory \cite{\morrii}. 

Importantly, notice that the reduction of parameters that takes place on trivialisation \eqref{fonelimit} from the infinitely many underlying couplings \eqref{coeffone} to the effective coupling $\kappa$ \eqref{kappa} (Newton's constant) and a cosmological constant \eqref{cc} at first order is \emph{not} the result of imposing infinitely many relations between these underlying couplings, but rather a dramatic demonstration of universality resulting from the large amplitude suppression scale limit. This reduction of parameters occurs provided only that the underlying couplings are chosen from some loosely specified infinite dimensional domain. Thus $f^1_\Lambda(\vp)$ is given in general by specifying its Fourier transform as \eqref{ffform} (with $n_\sigma\cu=0$, $A_\sigma\cu=\kappa$ and $\Lambda_\sigma\cu=\Lambda_\p$, as explained in sec. \ref{sec:first}). Similar remarks follow for $f^{1_1}_\Lambda$ following \eqref{ffforma}.
These Fourier transforms are proportional to the reduced Fourier transform $\bar{\ff}^\sigma(\bar{\vpi}^2)$. As we noted, at first order this latter function can be chosen to be independent of $\Lambda_\sigma$, then the only constraints on it,\footnote{At higher orders the only other constraints are the mild convergence conditions \eqref{vanishes}.} are that it is a dimensionless entire function, that it has asymptotic behaviour \eqref{largevpibar}  as $\bar{\vpi}\cu\to\infty$, and that its integral \eqref{normalised} is normalised. This still leaves an infinite dimensional function space. In particular any number of underlying couplings \eqref{fourier-expansion} can still take any value. A key result of this paper is the demonstration that the same results are then nevertheless recovered \cite{\morrii}, thus confirming universality. Indeed it is only the underlying couplings' asymptotic behaviour for large $n$ that is constrained through \eqref{largeg}, and it is only these values that ultimately influence the physical results, as discussed in deriving their uniform bound \eqref{gmax}.

This observation was also emphasised at the end of app. \ref{app:spectrum} when discussing coefficient functions with a spectrum of amplitude suppression scales. However in the body of the paper we recognised that we can make three simplifications to the most general case. As explained in sec. \ref{sec:general}, firstly we can work only with coefficient functions of definite parity, \ie even or odd under $\vp\cu\mapsto-\vp$, and secondly with coefficient functions containing only one amplitude suppression scale. Finally in sec. \ref{sec:first}, we also recognised that we can set all amplitude suppression scales to a common value $\Lambda_\sigma\cu=\Lambda_\p$. This still leaves us  to choose, for each coefficient function, a reduced Fourier transform function $\bar{\ff}^\sigma(\bar{\vpi}^2)$ with its own domain of infinitely many underlying couplings, and thus is more than sufficient again to demonstrate universality of the continuum and large amplitude suppression scale limits.

As we have seen, in the end at first order we are left with just the two effective couplings, Newton's constant and the cosmological constant. A key question \cite{\morri,\morrii} is  how many (effective) couplings are left once higher order quantum corrections are included. After all, it is at this point operationally, that one meets in standard perturbative quantisation an apparent obstruction to defining quantum gravity since new couplings get introduced to absorb divergences, order by order in perturbation theory. Given the importance of this question, we finish by commenting on this, although we cannot do better  than make some  remarks, since substantiation requires developments that go well beyond what we report in this paper. Although in refs. \cite{secondconf, second} we will establish that this continuum limit can be extended to second order for pure quantum gravity, this does not yet seem enough to settle the above question since, although we find that the new divergences can be absorbed by wave-function-like renormalization, this is also famously true of pure quantum gravity in its standard quantisation at this order \cite{tHooft:1974toh}. \textit{A priori} in this quantisation a continuum limit with an infinite number of couplings seems logically consistent \cite{second}. However as we will show, there are indications that the quantisation is more restrictive at higher orders, where the underlying couplings introduced here becoming running couplings  \cite{secondconf,second}. In particular, note again that so far we have been relying on the fact that the renormalized trajectory can be constructed in the $\vp$-sector by flowing upwards from the IR to the UV. At the linearised level this was set out precisely, together with its proof, at the end of sec. \ref{sec:newquantisation}. At higher orders this kind of `reverse' flow construction is also key \cite{secondconf}. However the flow in the $h_{\mu\nu}$ (graviton) sector is guaranteed only in the usual direction from the UV to the IR. Put together we are actually dealing with a flow equation that does not have a well-defined Cauchy initial value problem in either direction. In other words, a generic `initial' effective action will lead to singular flows in \emph{both} directions. This does not mean that there are no solutions (after all we just established one to first order here) but  we find \cite{second} that it does appear at higher orders to require solutions to depend ultimately on only the two parameters, Newton's constant and the cosmological constant.


\section*{Acknowledgments}
AM acknowledges support via an STFC PhD studentship. TRM thanks David Turton for discussions, and acknowledges support from STFC through Consolidated Grant ST/P000711/1. 

\appendix

\section{Further examples of coefficient functions}
\label{app:spectrum}
\vskip-10pt
\subsection{Examples with multiple amplitude suppression scales}
\vskip-5pt
Here we develop some of the properties of linearised coefficient functions that are constructed from a spectrum of amplitude suppression scales $\gamma_k\Lambda_\sigma$. For example for symmetric coefficient functions satisfying flat trivialisation \eqref{flat},
we can take \cite{\morrii} 
\be 
\label{ffsum}
\ff^\sigma(\vpi) = A_\sigma\sum_{k=0}^N a_k\, \ff(\vpi,\gamma_k\Lambda_\sigma)\,,
\ee 
where $N\ge \lceil \frac{n_\sigma}2\rceil$ will allow us to ensure that couplings $g^\sigma_{2n<n_\sigma}$ vanish,
and we define the function
\be 
\label{ffflatE}
\ff(\vpi,\bar\Lambda) = \sqrt{\pi}\, \bar\Lambda \,{\rm e}^{-\vpi^2\bar\Lambda^2/4}\,,
\ee
which is just the simplest choice of reduced Fourier transform \eqref{ffoneEg}, where for convenience we have absorbed the factor of $2\pi\Lambda_\sigma$ from the example  \eqref{ffsigmaEg}. 
The dimensionless parameters $\gamma_k>0$ are chosen unequal, and without loss of generality we order them and set the greatest to unity:
\be 
\label{gammak}
0<\gamma_N<\gamma_{N-1}<\cdots<\gamma_0=1\,,
\ee
and the dimensionless coefficients $a_k$ are chosen to satisfy 
\beal 
\label{ak1}
\sum_{k=0}^N a_k &= 1\,,\qquad\text{and}\\
\label{ak2}
\sum_{k=0}^N a_k\,\gamma_k^{2n+1} &= 0\qquad\ \ \text{for}\quad 0\le n< \left\lceil \frac{n_\sigma}2\right\rceil\,.
\eeal
Performing the integral in the Fourier transform representation \eqref{fourier-sol} we get
\be 
\label{fsum}
f^\sigma_\Lambda(\vp) = \sum_{k=0}^N a_k\,f_\Lambda(\vp,\gamma_k\Lambda_\sigma)\,,
\ee
where $f_\Lambda(\vp,\gamma_k\Lambda_\sigma)$ is just the $\alpha\cu=1$ example \eqref{foneEg} with $\Lambda_\sigma$ rescaled by $\gamma_k$.

From the definition of the amplitude suppression scale, see above \eqref{asympfLp}, we see that $f^\sigma_\Lambda$ has overall amplitude suppression scale $\Lambda_\sigma$, corresponding to the maximum one $\gamma_0\Lambda_\sigma =\Lambda_\sigma$. We verify that it also characterises the exponential decay of the physical coefficient function: setting $\Lambda=0$, 
\be 
\label{fsumphys}
f^\sigma\!(\vp) = A_\sigma\sum_{k=0}^N a_k\,\mathrm{e}^{-\vp^2/\gamma^2_k\Lambda_\sigma^2} \sim\, a_0\, A_\sigma \, \mathrm{e}^{-\vp^2/\Lambda_\sigma^2}\,,
\ee
where the last equation holds at large $\vp$. Thus we satisfy the asymptotic formula for the physical coefficient function \eqref{largephys}, but we have here an example where the asymptotic behaviour is fixed by $A_\sigma$ only up to an undetermined dimensionless proportionality constant, as already commented below \eqref{dimA}. Importantly note that the large $\vpi$ behaviour in the sum over a spectrum of amplitude suppression scales \eqref{ffsum} is however set by the smallest amplitude suppression scale:
\be 
\ff^\sigma(\vpi) \sim \sqrt{\pi}\,a_N \,\gamma_N\, A_\sigma \, \Lambda_\sigma\, \,{\rm e}^{-\vpi^2\gamma^2_N\Lambda^2_\sigma/4}\,,
\ee
and thus the asymptotic formula for the Fourier transform \eqref{largeFourier} does not hold, hence the comments below it.
 The couplings in the Taylor expansion of the Fourier transform \eqref{fourier-expansion} are given by
\be 
\label{gseven}
g^\sigma_{2n} = \frac{\sqrt{\pi}}{n!4^n}\,A_\sigma\Lambda_\sigma^{2n+1} \sum_{k=0}^N a_k\,\gamma_k^{2n+1}\sim\, a_0\, A_\sigma \frac{\sqrt{\pi}}{n!4^n}\,\Lambda_\sigma^{2n+1} \,,
\ee
and satisfy the constraint that they vanish for $2n<n_\sigma$, thanks to the vanishing summation constraint \eqref{ak2}. The last equation holds at large $n$, which thus verifies that the asymptotic formula for couplings \eqref{largeg} nevertheless holds, although again we see the presence of an undetermined proportionality. 
Finally, since $f_\Lambda(\vp,\gamma_k\Lambda_\sigma)\to1$ as $\Lambda_\sigma\to\infty$, we have from the sum normalisation constraint \eqref{ak1} that flat trivialisation \eqref{flat} is satisfied, while since $\sqrt{\vphantom{\tilde\Lambda}\Lambda^2+a^2\gamma^2_k\Lambda^2_\sigma} $ sets the scale for $\vp$-variation in the components, we see that the flat limit \eqref{flat} is reached at least as fast as $O(1/\gamma_N\Lambda_\sigma)$ and more generally the refined limit
\eqref{regularity} is satisfied. Notice however that it is the smallest amplitude suppression scale that controls the corrections here.

 Since the summation constraints  (\ref{ak1},\ref{ak2}) provide $\lceil \frac{n_\sigma}2\rceil+1$ linearly independent conditions on $N+1\ge \lceil \frac{n_\sigma}2\rceil+1$ coefficients $a_k$, they can always be satisfied. By choosing $N>\lceil \frac{n_\sigma}2\rceil$ large enough, we can go on to fix the numerical coefficient of finitely many of any of the surviving $g^\sigma_{2n}$ (with $n$ finite) to any value we wish, including forcing them also to vanish. We also have the freedom to alter couplings through changing the $0<\gamma_{k>0}<1$ provided they remain unequal. We see 
that the flat trivialisation limit \eqref{flat} is independent of the value of any finite set of finite-$n$ couplings or indeed of any finite number of relations between these couplings \cite{secondconf}. Therefore, apart from confirming that we can ensure that $g^\sigma_{2n<n_\sigma}=0$, the universal information on the couplings is that captured in the large $n$ asymptotic estimate \eqref{largeg}, which indeed holds for any linearised solution.

For examples satisfying polynomial trivialisation \eqref{flatp}, we can still use the sum over a spectrum of amplitude suppression scales \eqref{ffsum}, where by the map to a Fourier transform for a coefficient function satisfying the polynomial trivialisation constraint \eqref{ffsigmap}, we replace $\ff(\vpi,\bar\Lambda)$ with $\left( i \partial_\vpi\right)^\alpha \ff(\vpi,\bar\Lambda)$ along the lines already discussed in sec. \ref{sec:examples}.

\subsection{Other examples with only one amplitude suppression scale}
\vskip-5pt
As explained in sec. \ref{sec:general} we insist in this paper on using only one amplitude suppression scale, and our examples are all expressible in conjugate momentum space as an exponential decay factor times a polynomial as in secs. \ref{sec:examples}. 
Other examples with only one amplitude suppression scale could be generated, \eg 
\be 
\label{ffsumalt}
\ff^\sigma(\vpi) = A_\sigma\sum_{k=0}^N a_k\, \ff(\vpi,\gamma_k,\Lambda_\sigma)\,,
\ee 
for appropriate choices of $a_k$, where we choose the function to be
\be 
\ff(\vpi,\gamma,\bar\Lambda) = \sqrt{\pi}\, \bar\Lambda \,{\rm e}^{-\left(\vpi^2\bar\Lambda^2+\gamma^2\right)/4} \cosh(\gamma\bar\Lambda\vpi/2)\,,
\ee
corresponding to the physical coefficient function
\be 
f\!(\vp,\gamma,\bar\Lambda) = \mathrm{e}^{-\vp^2/\bar\Lambda^2}\cos(\gamma\vp/\bar\Lambda)\,,
\ee
which thus gives the $\Lambda>0$ solution
\be 
f_\Lambda(\vp,\gamma,\bar\Lambda) =   \frac{ a\bar\Lambda}{\sqrt{\vphantom{\tilde\Lambda}\Lambda^2+a^2\bar\Lambda^2}}\, \exp\left(-\frac{a^2\vp^2+\gamma^2\Lambda^2/4}{\vphantom{\tilde\Lambda}\Lambda^2+a^2\bar\Lambda^2}\right)\,\cos\left(\frac{a^2\gamma\bar\Lambda\vp}{\vphantom{\tilde\Lambda}\Lambda^2+a^2\bar\Lambda^2}\right) \,,
\ee
which clearly again has the right limiting properties to satisfy flat trivialisation \eqref{flat} and the refined limits \eqref{regularity}.
These functions have the same amplitude suppression scale $\bar\Lambda$ irrespective of the choice of $\gamma$.
Further examples can be generated by exchanging $\cosh$ with $\cos$ in the above, or for odd functions, replacing these with $\sinh$ and sine. 

%


\bibliographystyle{hunsrt}
\bibliography{references} 

\end{document}